\documentclass[twocolumn,twoside]{IEEEtran}

\ifCLASSINFOpdf
  
\else
 
\fi

\ifCLASSOPTIONcompsoc
\usepackage[caption=false, font=normalsize, labelfont=sf, textfont=sf]{subfig}
\else
\usepackage[caption=false, font=normalsize]{subfig}
\fi
\usepackage{lipsum}%
\usepackage[dvipsnames]{xcolor}

\usepackage{balance}
\usepackage{multicol}   % for equalize of last page columns
\usepackage{cite}
\usepackage{gensymb}
\usepackage{multirow}
\usepackage{graphics}
\usepackage{epsfig}
\usepackage{graphicx}
\usepackage{epstopdf}
\usepackage{textcomp}
\usepackage{amsmath}
\usepackage{mathtools}
\usepackage{filecontents}
\usepackage{lipsum,color}
\usepackage{amssymb}
\usepackage{float}
\usepackage{times} % assumes new font selection scheme installed
\usepackage{amsthm}  % for proof 
\usepackage{amsfonts}

\theoremstyle{break}

\begin{document}
%\title{Optical Channel Modeling Between UAVs over Weak to Strong Turbulence Channels}
%\title{Performance Analysis of UAV to Ground Optical Wireless Link based on Modulating Retro-Reflector }
\title{Modulating Retroreflector Based Free Space Optical Link for UAV-to-Ground Communications}

%\author{Author A and Author B
\author{M.~T.~Dabiri,~M.~Rezaee,~L.~Mohammadi,~F.~Javaherian,~V.~Yazdanian,\\~M.~O.~Hasna,~{\it Senior Member,~IEEE},~and~M.~Uysal,~{\it Fellow,~IEEE}}

\maketitle

%%%%%%%%%%%%%%%%%%%%%%%%%%%%%%%%%%%%%%%%%%%%%%%%%%%%%%%%%%
%%%%%%%%%%%%%%%%%%%%%%%%%%%%%%%%%%%%%%%%%%%%%%%%%%%%%%%%%%
\begin{abstract}
%%%%%%%%%%%%%%%%%%%%%%%%%%%%%%%%%%%%%%%%%%%%%%%%%%%%%%%%%%
%%%%%%%%%%%%%%%%%%%%%%%%%%%%%%%%%%%%%%%%%%%%%%%%%%%%%%%%%%
Weight reduction and low power consumption are key requirements in the next generation of unmanned aerial vehicle (UAV) networks.
Employing modulating retro-reflector (MRR)-based free space optical (FSO) technology is an innovative technique for UAV-to-ground communication in order to reduce the payload weight and power consumption of UAVs which leads to increased maneuverability and flight time of UAV.
In this paper, we consider an MRR-based FSO system for UAV-to-ground communication.
We will show that the performance of the considered system is very sensitive to tracking  errors.
Therefore, to assess the benefits of MRR-based UAV deployment for FSO communications, the MRR-based UAV FSO channel is characterized by taking into account tracking system errors along with UAV's orientation fluctuations, link length, UAV's height, optical beam divergence angle, effective area of MRR, atmospheric turbulence and optical channel loss in the double-pass channels.
To enable effective performance analysis, tractable and closed-form expressions are derived for probability density function of end-to-end signal to noise ratio, outage probability and bit error rate of the considered system under both weak-to-moderate and moderate-to-strong  atmospheric turbulence conditions. The accuracy of the analytical expressions is verified by extensive simulations.
Analytical results are then used to study the relationship between the optimal system design and tracking system errors.
\end{abstract}

\begin{IEEEkeywords}
Angle of arrival (AoA) fluctuations, FSO communications, UAV, modulating retro-reflector (MRR).
\end{IEEEkeywords}
\IEEEpeerreviewmaketitle
%%%%%%%%%%%%%%%%%%%%%%%%%%%%%%%%%%%%%%%%%%%%%%%%%%%%%%%%%%%%
%%%%%%%%%%%%%%%%%%%%%%%%%%%%%%%%%%%%%%%%%%%%%%%%%%%%%%%%%%%%
\section{Introduction}
%%%%%%%%%%%%%%%%%%%%%%%%%%%%%%%%%%%%%%%%%%%%%%%%%%%%%%%%%%%%
%%%%%%%%%%%%%%%%%%%%%%%%%%%%%%%%%%%%%%%%%%%%%%%%%%%%%%%%%%%%
%%%%%%%%%%%%%%%%%%%%%%%%%%%%%%%%
%\subsection{Background}
%%%%%%%%%%%%%%%%%%%%%%%%%%%%%%%%
\IEEEPARstart{U}{nmanned} aerial vehicles (UAVs) have been introduced to overcome many of the shortcomings of the current terrestrial infrastructure by operating as aerial communication nodes and providing robust line-of-sight (LoS) connectivity to ground devices \cite{mozaffari2019tutorial,dabiri2020analytical}.
%easily deployed and agile enough to maneuver and adjust their altitudes depending on conditions.
% makes them promising candidates for a wide range of applications.
%is a flexible and cost-effective approach to providing on-demand communications [2]–[
UAVs deployed as flying communication nodes using radio frequency (RF) will interfere with ground devices, hence degrading the performance of the ground network.
Employing UAVs equipped with free space optical (FSO) technology is a promising method for future ultra dense wireless networks \cite{alzenad2018fso}.
FSO communication systems use very small beam divergence which is physically inaccessible to RF technologies, thus, making FSO links extremely secure.
Moreover, FSO system is much faster, easier to deploy, more compact, and cheaper than RF \cite{mynaric,khalighi2014survey,kaushal2016optical}.
However, vulnerability to signal blockage is one of the fundamental limitations of FSO links that essentially confines the receiver to be placed within the line-of-sight (LoS) of the transmitter. A potential application of UAV-assisted FSO systems is in dense cities with tall buildings where a UAV can act as an aerial relay to connect source and destination nodes where the LoS between ground optical nodes is interrupted by tall buildings \cite{dabiri2019optimal,9492795}.

%------------------------------
%------------------------------
\subsection{Literature Review and Statement of the Problem}
Recently, significant research works have been focused on communication problems of UAV-based FSO systems and how to utilize their vulnerabilities \cite{fawaz2018uav,                zhang20203d,     
	gu2020topology,gu2020optimizing,                 lee2019uav,                 ajam2020ergodic,najafi2020statistical,
	dabiri2018channel,dabiri2019tractable, dabiri2020channel
     }. 
%--------------------------- 
For instance, two possible scenarios are proposed in \cite{fawaz2018uav} for the integration of UAVs as buffer-aided moving relays into the conventional relay-assisted FSO systems.
In \cite{zhang20203d}, the 3D deployment and resource allocation of a UAV Base Station with FSO-based backhaul is studied in a given hotspot area. 
In \cite{gu2020topology,gu2020optimizing}, the authors design an efficient algorithm for FSO-based UAVs relay network topology to achieve a high network reliability.
The trajectory optimization of a fixed-wing UAV using FSO communication is addressed in \cite{lee2019uav}.
In particular, the authors focus on maximizing the flight time of the UAV by considering practical constraints including limited propulsion energy and required data rates. 
The very small beam divergence which inherently increases the secrecy and capacity of an FSO link  makes it very sensitive to beam misalignment. 
Unlike an stable ground node, the position and orientation of UAVs fluctuate due to independent random effects such as wind speed, changes in the air pressure, propeller rotation, engine operation, attitude control system faults and platform stability error \cite{dabiri20203d}. 
However, the results of \cite{fawaz2018uav,                zhang20203d,     
	gu2020topology,gu2020optimizing,                 lee2019uav}  
are obtained by neglecting the effect of UAVs' random fluctuations which are valid for larger UAVs equipped by  fast and precise stabilizers. % which is bulky and very expensive. 

More recently, the authors in \cite{ajam2020ergodic,najafi2020statistical,
	dabiri2018channel,dabiri2019tractable, dabiri2020channel}, 
studied the problem of UAV-based FSO links by including the effects of UAV's orientation and position fluctuations.
In \cite{ajam2020ergodic}, the authors analyzed the end-to-end system performance of mixed RF-FSO networks employing UAVs as buffer-aided 
and non-buffer-aided relays in terms of the ergodic sum rate by taking into account the impacts of UAV's fluctuations.
A novel FSO channel model between a hovering UAV and a central unit is developed in \cite{najafi2020statistical} by quantifying the corresponding geometric and misalignment losses, while taking into account the non-orthogonality of the laser beam. 
Considering the joint effects of UAVs’ fluctuations as well as atmospheric turbulence, a novel channel model was proposed in \cite{dabiri2018channel,dabiri2019tractable} that is suitable for hovering UAV-based FSO links with zero boresight angle.
In \cite{dabiri2020channel}, the authors completed the results of \cite{dabiri2018channel,dabiri2019tractable} and provided a novel UAV-based FSO channel model that takes into account the effect of nonzero boresight pointing errors.
In \cite{najafi2020statistical}, the considered value for standard deviation (SD) of orientation fluctuations is $\sigma_{\theta_o}=0.2-0.8$\,mrad (or equivalently is  $\sigma_{\theta_o}=0.0115^o-0.0458^o$ in degrees), in \cite{dabiri2019tractable} is  $\sigma_{\theta_o}=5$\,mrad, and in \cite{dabiri2020channel} is $\sigma_{\theta_o}=1-5$\,mrad.
However, due to the flight time, payload,  and power consumption limitations of lightweight multi-rotor UAVs, reaching such values for SD of orientation fluctuations may not be always possible.
Another challenge is an asymmetric nature of the ground-to-UAV and the UAV-to-ground links.
Although the ground-to-UAV link can withstand larger orientation fluctuations, the results of \cite{dabiri2018channel,dabiri2019optimal} show that severe fluctuations of UAVs greatly reduce the quality of the UAV-to-ground link. 
The ground station (GS) has more power and payload capacity than the UAV node and employing a fast and precise tracking system in GS node is feasible. Therefore, it can well track the UAV node.
In a ground-to-UAV link, the transmitter (Tx) is mounted on an stable GS node with precise tracking system, thus,
by increasing UAV's orientation fluctuations,  only the SD of angle-of-arrival (AoA) increases in a ground-to-UAV link and the receiver's (Rx's) field-of-view (FoV) can be increased to relax this degrading effect \cite{safi2020analytical}.
However, in a UAV-to-ground link, the Tx is mounted on an unstable UAV. For better understanding, consider a 1000\,m UAV-to-ground link.
When the orientation of Tx deviates more than 10\,mrad, the received optical beam deviates more than 10\,m from the center of Rx's aperture, and this leads to an unreliable communication.
Moreover, for compensation of UAV's orientation fluctuations in a UAV-to-ground link, the Tx needs a power amplifier to increase the transmitted power which increases the payload and power consumption of UAV and limits the maneuverability and flight time of UAV.

Modulating retro-reflectors (MRR)\footnote{An MRR consists of an optical retroreflector with a modulator to first modulate the incoming optical signal and then reflect it toward the transmitter. This feature makes it possible to act as an optical communications device without sending its own optical power as graphically depicted in Fig. \ref{r2}. A number of technologies have been considered for the modulation component, including electro-optic modulators, liquid crystal modulators, multiple quantum well devices, and actuated micromirrors. The modulator tries to block the reflected signal intensity for a bit "0" and tries to pass all the reflected signals for a bit "1". In other words, the modulator changes the intensity of the reflected optical signal in proportion to the transmitted On–off keying (OOK) signal sequence, which is called switching speed or switching rate.} is potentially attractive in asymmetric situations such as small UAV platforms which are too small to carry a conventional FSO terminal \cite{goetz2012modulating}.\footnote{Here, the GS has more power and payload capacity than the UAV node, and GS sends light towards the remote Rx mounted on UAV. The UAV is equipped with a small MRR which upon sensing the incoming interrogating beam, modulates and reflects it directly back to the GS \cite{kaushal2016underwater}.}
Retro-reflector links are used in limited duplex communication where Rx have low power to support full transceiver operations. 
MRRs are also used to reduce the pointing and tracking requirements by directly reflecting the incoming light to the GS independent of orientation of the retro-reflector \cite{kaushal2016underwater}.
In the MRR-based topology, complexity and tracking equipment are transferred from the UAV node to the GS. 
Even though MRR-based optical wireless communications have been well studied in the context of underwater optical wireless communications \cite{kaushal2016underwater}, this subject is restricted to few works in the context of UAV-based FSO communications \cite{quintana2016design,quintana2014novel,     
	yang2017wave,yang2017channel,yang2018performance,yang2020wavefront,     
    li2017probability,li2019bit}. 
In \cite{quintana2016design,quintana2014novel}, the authors design and implement a real-time localization and tracking system for a UAV-to-ground FSO link.
In \cite{yang2017wave}, a general geometrical model of the corner cube reflector (CCR) is established based on the ray tracing method and then, the authors used the  Wave Optics simulations to investigate the double-pass channel in the MRR-based FSO systems.
The probability density function (PDF) and cumulative distribution function (CDF) of the double-pass MRR-based FSO systems
are derived  in \cite{yang2017channel} under weak turbulence conditions and in \cite{yang2018performance} under strong turbulence conditions.
Impacts of shape and size of MRR cell along with turbulence condition on the beam spot are investigated in \cite{yang2020wavefront} by the theoretical analysis and wave-optics simulation. 
In \cite{li2017probability,li2019bit}, the effects of the parameters such as atmospheric turbulence conditions, link length, Rx's aperture diameter and the average signal-to-noise ratio (SNR) are studied on the performance of MRR-based FSO links.
However, the results of \cite{yang2017wave,yang2017channel,yang2018performance,yang2020wavefront,     
	li2017probability,li2019bit}  
are obtained by neglecting the effect of geometrical pointing errors which are only valid when the GS node perfectly pointed optical beam towards the aperture of MMR mounted on UAV.

Even though a larger value of MRR's aperture improves the link budget, in a practical implementation, we are not allowed to use large values for MRR's aperture because the switching rate of MRR modulator is inversely proportional to MRR's aperture. 
This leads to a large geometrical loss in an MRR-based FSO system with respect to the conventional FSO systems with much larger aperture area. 
To compensate  this problem, the beamwidth must be chosen much smaller than the values of the beamwidth used in conventional FSO systems. 
This makes the MRR-based FSO system very sensitive to tracking errors. 
Depending on link length, we will show that any tracking angle error in the order of $\micro$rad can significantly affect the performance of MRR-based FSO systems. 
Therefore, to assess the benefits of MRR-based UAV deployment for FSO communications,  performance analyses of the considered system under tracking system errors is very important and necessary.
To the best of authors' knowledge, there is no prior work in the literature that models and analyzes MRR-based FSO systems for UAVs under tracking system errors by taking into account the effects of MRR's orientation fluctuations as well as atmospheric turbulence conditions. 
%

%
%%%%%%%%%%%%%%%%%%%%%%%%%%%%%%%%%%%%%%%%%%%%%%%%%%%%%%%%%%%%%%%%
%%%%%%%%%%%%%%%%%%%%%%%%%%%%%%%%%%%%%%%%%%%%%%%%%%%%%%%%%%%%%%%% VERSUS W_Z
\begin{figure}
	\centering
	\subfloat[] {\includegraphics[width=3.3 in]
		{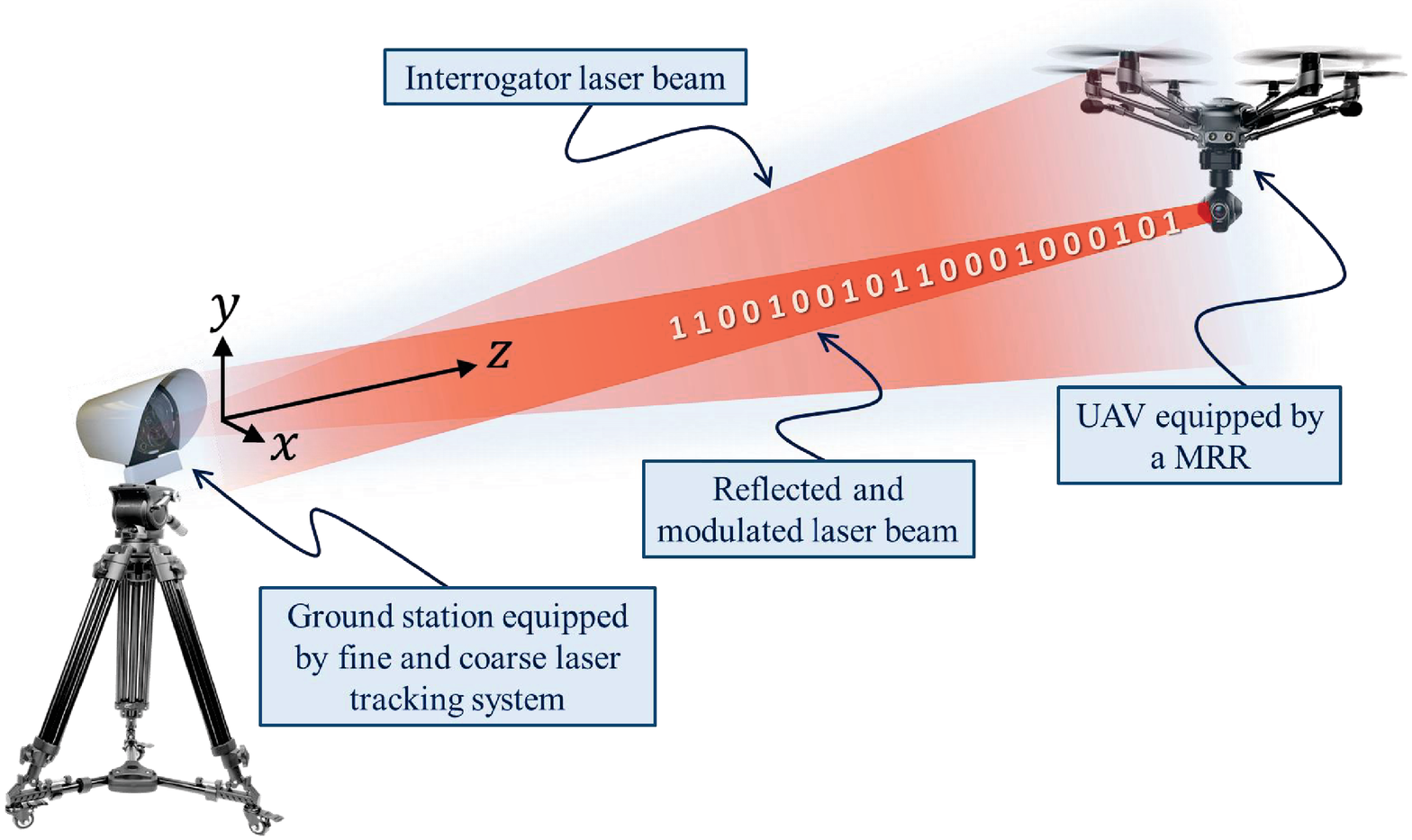}
		\label{r1}
	}
	\hfill
	\subfloat[] {\includegraphics[width=3.1 in]{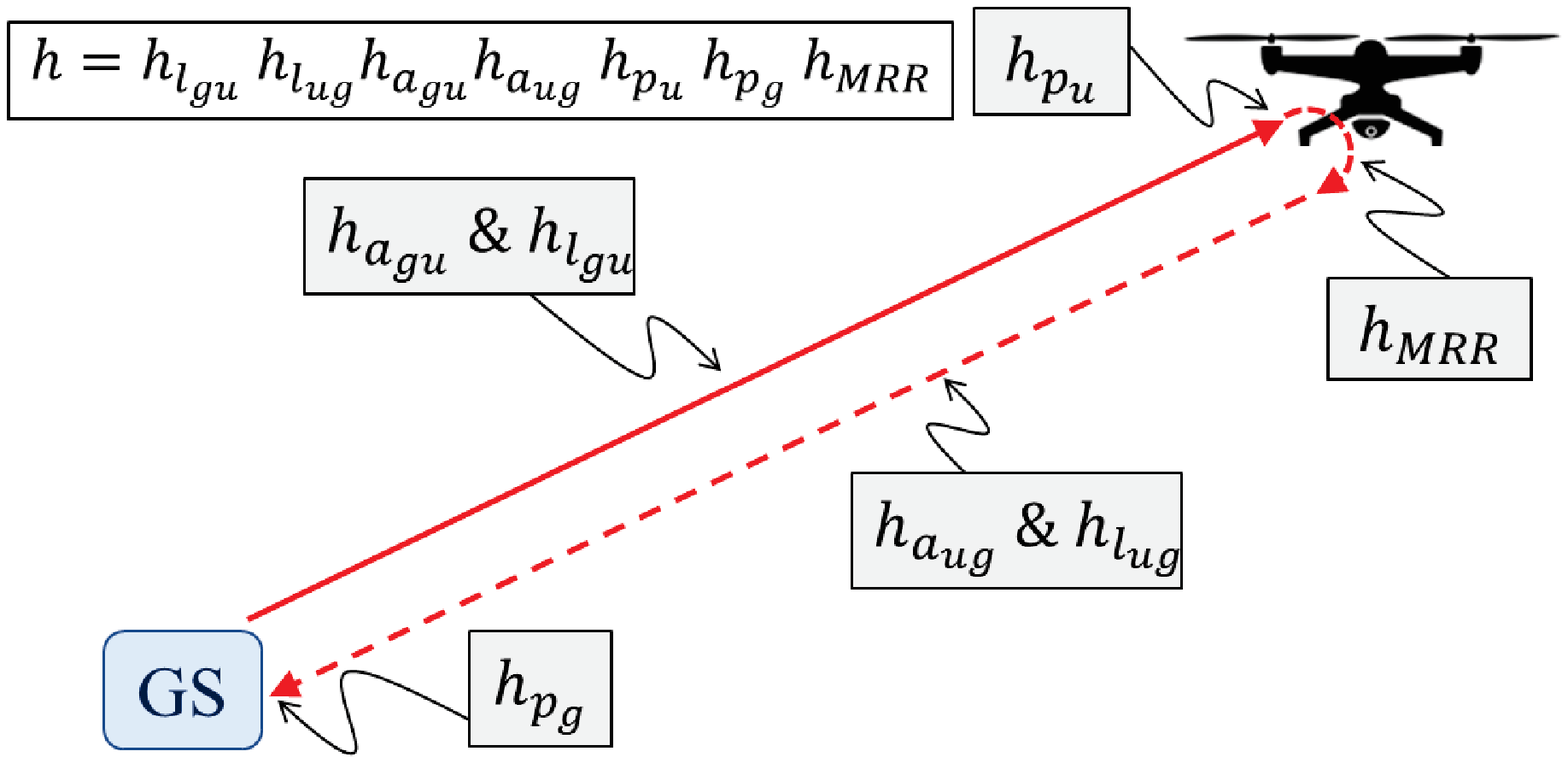}
		\label{r3}
	}
	\caption{The schematic of the considered communication link between an optical GS and an hovering UAV equipped with an MRR. The $z$ axis refers to the direction that extends from GS toward UAV node and the GS tries to adjust optical beam in the direction of $z$-axis. MRR mounted on UAV modulates and reflects the incoming beam directly back to the GS. 
	(b) Our GS-to-UAV-to-GS channel model takes into account seven impairments.
	}
	\label{cxd}
\end{figure}
%%%%%%%%%%%%%%%%%%%%%%%%%%%%%%%%%%%%%%%%%%%%%%%%%%%%%%%%%%%%%%%%
%%%%%%%%%%%%%%%%%%%%%%%%%%%%%%%%%%%%%%%%%%%%%%%%%%%%%%%%%%%%%%%%  
%

%---------------------------------------
%---------------------------------------
\subsection{Major Contributions and Novelty}
%---------------------------------------
%
In this paper, we consider the MRR-based FSO system for  UAV-to-ground communication  where the GS sends an unmodulated continuous laser beam towards the MRR mounted on UAV. The incident beam is modulated by the MRR and is directly reflected back to the GS.
The main contribution of this paper is the performance analysis and system design of the considered UAV-based FSO system when UAV is equipped with MRR under tracking errors by taking into account the effects of MRR's orientation fluctuations as well as atmospheric turbulence conditions.  
In summary, our key contributions include:

\begin{itemize}
	\item We develop channel models for the UAV-to-ground  MRR-based FSO communication system for both weak-to-moderate and moderate-to-strong  atmospheric turbulence conditions, 
	by taking into account  
	tracking errors, UAV's orientation fluctuations, link length, UAV's height, optical beam divergence angle, effective area of MRR, atmospheric turbulence and optical channel loss in the double-pass channels.
	%%%%%%%%%%%%%%%%%%%%%%%
	\item Based on these models, we derive closed-form analytical expressions for PDF under both weak-to-moderate and moderate-to-strong  atmospheric turbulence conditions. 
	Then, through Monte Carlo simulations, the accuracy of the derived statistical distributions is verified.
	%%%%%%%%%%%%%%%%%%%%%%%
	\item We also derive the closed-form expressions for the PDF and CDF of end-to-end signal-to-noise ratio (SNR), outage probability and bit error rate (BER)  of the MRR-based UAV FSO system. The accuracy of the analytical expressions is verified by using simulations.
	Analytical results are then used to study the impact of the system parameters on the performance of  MRR-based UAV FSO system under different conditions, e.g., a wide range of tracking errors, atmospheric turbulence strengths, different levels of UAV's instability, different link lengths, etc.
	Our results reveal that unlike conventional FSO systems, optimal design of a UAV-based MRR FSO system is very important and any change in the parameters (such as link length, SD of tracking errors, target BER, desired data rate, etc.) affect the optimal values of other parameters. 
\end{itemize}

%---------------------------------------------
%---------------------------------------------
\subsection{Organization}
We list the main notations in Table \ref{I1}.
The organization of the rest of the paper is as follows. In Section II, we characterize the actual channel models of MRR-based UAV FSO system.
Then, in Section III, we provide the analytical channel models.
Next, in Section IV, we provide the simulation results to verify the derived analytical channel models and study the link performance and system parameter optimization.
Finally, conclusions are drawn in Section V.

%

%%%%%%%%%%%%%%%%%%%%%%%%%%%%%%%%%%%%%%%%%%%%%%%%%%%%%%%%%%%%%%%%
\begin{table}
	\caption{The list of main notations.} % title of Table
	\centering % used for centering table
	\begin{tabular}{l c} % centered columns (3 columns)
		\hline\hline \\[-1.2ex]%inserts double horizontal lines\\
		{\bf Parameter} & {\bf Description}  \\ [.5ex] % inserts table
		%heading
		\hline\hline \\[-1.2ex]% inserts single horizontal line
		$[x; y; z]$ & Cartesian coordinate system that  $z$ axis refers to the \\[-.5ex]
		& direction that extends from GS toward UAV node\\ [.5ex]
		$[x'; y'; z']$ & Cartesian coordinate system that indicates the \\[-.5ex]
		& coordinates of three perpendicular mirrors of MRR\\ [.5ex]
		$R$& PD responsivity    \\
		$\Upsilon_\textrm{th}$& SNR threshold   \\
		$r_g$ & Radius of GS aperture \\
		$ A_r $  &    MRR effective area               \\
		$ Z $      & Link length        \\
		$\sigma_{\theta_o}$ & SD of UAV orientation fluctuations   \\
		$\sigma_{\theta_e}$ & SD of tracking angle errors   \\
		$ \lambda $ &  Wavelength                         \\
		$\theta_\textrm{div}$  & Divergence angle   \\
		$w_z$ & Beamwidth at the Rx        \\
		$P_t$   & Transmit power                    \\ 
		$\sigma_n^2$   & Noise variance           \\
		$N$  & Number of sectors  \\
		$ h_{l_{gu}}$ & Channel Loss of GS to UAV link                    \\
		$ h_{l_{ug}}$ & Channel Loss UAV to GS link                    \\
		$ h_{a_{gu}}$ & Atmospheric turbulence coefficient of GS to UAV link                    \\
		$ h_{a_{ug}}$ & Atmospheric turbulence coefficient of UAV to GS link                    \\
		$h_{p_u}$ & Attenuation due to pointing errors at the MRR aperture   \\
		$h_{p_g}$ & Attenuation due to geometric loss in GS \\
		$h_{MRR}$ & The ratio of direct reflected power by MRR \\
		$C_n^2$ & Refractive-index structure  \\
		\hline \\[-1.2ex]
		$f_x(x)$ & The PDF of RV $x$ \\
		$F_x(x)$ & The CDF of RV $x$ \\
		$Q(\cdot)$ & The {\it Q}-function defined in \cite{jeffrey2007table} \\
		$\textrm{erf}(\cdot)$ & The error function defined in \cite{jeffrey2007table}\\
		$\textrm{erfc}(\cdot)$ & The complementary error function defined in \cite{jeffrey2007table}\\
		$G^{m,n}_{p,q}\!\!\left(\! z \Big|\!\!\!\!
		\begin{array}{c}
		-  \\
		-  
		\end{array}
		\!\!\!\right)$ &  The Meijer's G-function defined in \cite{wolfram2} \\
		\hline \hline              
	\end{tabular}
	\label{I1} % is used to refer this table in the text
\end{table}
%%%%%%%%%%%%%%%%%%%%%%%%%%%%%%%%%%%%%%%%%%%%%%%%%%
%%%%%%%%%%%%%%%%%%%%%%%%%%%%%%%%%%%%%%%%%%%%%%%%%%

%
%
%--------------------------------------------
\section{System Model}
%--------------------------------------------
%
%

%
%%%%%%%%%%%%%%%%%%%%%%%%%%%%%%%%%%%%%%%%%%%%%%%%%%%%%%%%%%%%%%%%
%%%%%%%%%%%%%%%%%%%%%%%%%%%%%%%%%%%%%%%%%%%%%%%%%%%%%%%%%%%%%%%%
\begin{figure}
	\begin{center}
		\includegraphics[width=3.3 in]{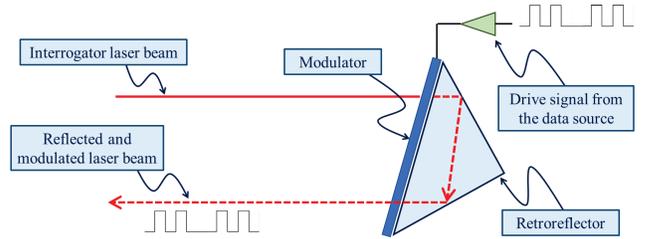}
		\caption{An MRR using a transmissive device is illustrated where the interrogation beam is modulated and directly reflected toward the incoming direction.}
		\label{r2}
	\end{center}
\end{figure}
\begin{figure}
	\begin{center}
		\includegraphics[width=1 \linewidth, draft=false]{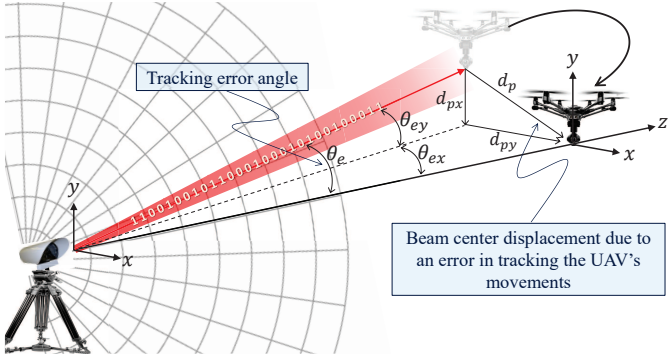}
		\caption{Graphical illustration of a ground-to-UAV FSO link wherein the ground station tracks the UAV mobility with the angle error $\sigma_{\theta_e}$.}
		\label{Ground-to-UAV}
	\end{center}
\end{figure}

%%%%%%%%%%%%%%%%%%%%%%%%%%%%%%%%%%%%%%%%%%%%%%%%%%%%%%%%%%%%%%%%
%%%%%%%%%%%%%%%%%%%%%%%%%%%%%%%%%%%%%%%%%%%%%%%%%%%%%%%%%%%%%%%%
%
%1860273335
Figure \ref{r1} shows MRR-based FSO link between an optical GS and an hovering UAV equipped with an MRR. We assume that the GS is located at $[0; 0; 0]$ (in Cartesian coordinate system $[x; y; z]$) and directs a continuous laser interrogator beam towards the UAV (located at $[0; 0; Z]$), in which $Z$ is the link length between GS and UAV. 
As illustrated in Fig. \ref{r2}, MRR modulates and reflects the incoming beam back to the GS.
MRR is suitable for the small UAVs with low power and payload limitations. %that is not possible to employ high quality antenna stabilizers and laser power amplifier due to the payload and power limitations.
It is used to reduce the pointing and tracking requirements by retro-reflecting  the modulated light back to the interrogating source. 
%
%On the other hand, the ground BS has no weight and power limits.
%
%
On the other hand, the GS is equipped with a precise laser tracking system that points a continuous laser interrogation beam towards the UAV. %In practice, every tracking system comes with a random error.
The accuracy of a tracking system is evaluated by the SD of angle errors. %, $\sigma_{\theta_e}$.
As depicted in Fig. \ref{Ground-to-UAV}, let $\theta_{ex}\sim \mathcal{N}(0,\sigma_{\theta_e}^2)$ and $\theta_{ey}\sim \mathcal{N}(0,\sigma_{\theta_e}^2)$ denote tracking system angle errors in the directions of $x$ and $y$ axes, respectively, with $\sigma_{\theta_e}$ being the SD of the angle errors. These tracking errors cause a radial distance between the received beam center and the MRR aperture center as $d_p=\sqrt{d_{px}^2+d_{py}^2}$, where $d_{px}$ and $d_{py}$ are the distance in the directions of $x$ and $y$ axes, respectively. 
Based on Fig. \ref{Ground-to-UAV}, any tracking error in the direction of $x$ and $y$ is formulated as
\begin{align}
	\label{pon1}
	\sin\left(\theta_{ex}\right) = \frac{d_{px}}{ \hat{Z}+z_e},	~~~ \&~~~ \sin\left(\theta_{ey}\right) = \frac{d_{py}}{ \hat{Z}+z_e}.
\end{align}
where $z_e=Z-\hat{Z}$, and $\hat{Z}$ is the estimated value of $Z$. 
In practice, $Z_e$ is in the order of a few tens of cm, however, $Z$ is in the order of a few hundred meters to a few km ($Z_e<<Z$), and thus, with a good accuracy, we can approximate \eqref{pon1} as
\begin{align}
	\label{po1}
	\sin\left(\theta_{ex}\right) = \frac{d_{px}}{ Z}\simeq\frac{d_{px}}{ \hat{Z}},	~~~ \&~~~ \sin\left(\theta_{ey}\right) =\frac{d_{py}}{ Z}\simeq \frac{d_{py}}{ \hat{Z}}.
\end{align}
We consider a Gaussian beam at the GS, for which the normalized spatial distribution of the received intensity at distance $Z$, is given by \cite{saleh2019fundamentals}
\begin{align}
I_r(d,Z) = \frac{2}{\pi w_z^2} \exp\left(-\frac{2(x^2 + y^2)}{w_z^2} \right),
\end{align}
where $d = [x, y]$ is the radial distance vector from the beam center. Also, $w_z$ is the beamwidth at distance $Z$ and can be approximated as $w_z\simeq \theta_{div} Z$, where $\theta_\textrm{div}$ is the optical beam divergence angle \cite{ghassemlooy2019optical}. %\cite{ricklin2002atmospheric}
%\begin{align}
%\label{ss1}
%w_z\simeq w_0 \sqrt{1+\epsilon \frac{\lambda Z}{\pi w_0^2} },
%\end{align}
%where $w_0$ is the beamwidth at $Z=0$, $\epsilon=\left(1+2w_0^2/\rho^2(Z)\right)$, $\rho(Z)=(0.55 C_n^2 k^2 Z)^{-3/5}$ is the coherence length, $C_n^2$  is the index of refraction structure parameter, $k=2\pi/\lambda$ is the optical wave number, and $\lambda$ is the optical wavelength. For small values of $w_0$, \eqref{ss1} 
%can be well simplified as $w_z\simeq \theta_{div} Z$, where $\theta_\textrm{div}$ is the optical beam divergence angle \cite{ghassemlooy2019optical}.

%wo is the beam waist at z = 0, ε = (1 + 2wo2/ρ2 o(z)), and ρo(z) = (0.55Cn2k2z)−3/5 is the coherence length.

\begin{figure}
	\begin{center}
		\includegraphics[width=2.3 in]{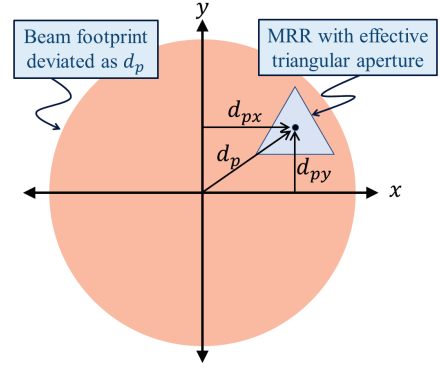}
		\caption{Gaussian beam footprint at the MRR with the effective triangular aperture area. The tracking errors cause a radial distance between the received beam center and the MRR aperture center as $d_p=\sqrt{d_{px}^2+d_{py}^2}$, where $d_{px}$ and $d_{py}$ are the distance in the directions of $x$ and $y$ axes, respectively.}
		\label{r4}
	\end{center}
\end{figure}

%
%  consists of three perpendicular triangular mirrors that are located in Coordinate planes $x'-y'$, $x'-z'$, and $y'-z'$, as depicted in Fig. \ref{cx1} where $[x'; y'; z']$ is a new Cartesian coordinate system. 
The considered MRR consists of three perpendicular triangular mirrors which has an aperture with a triangular effective area as depicted in Figs. \ref{r4} and \ref{cx1}.
%
%The aperture of MRR has a triangular effective area. 
For considered MRR-based topology with effective area $A_r$, the effective channel coefficient due to geometric spread with pointing error $d_p=\sqrt{d_{px}^2+d_{py}^2}$ is obtained as \cite{ghassemlooy2019optical}
\begin{align}
\label{ss2}
h_{p_u} &= \frac{2}{\pi w_z^2} \int \int_{p_{A_r}(x,y)}	 \nonumber \\
&~~~\times \exp\left(-\frac{2((x-d_{px})^2 + (y-d_{py})^2)}{w_z^2} \right)  \textrm{d}x \textrm{d}y,
\end{align}
where $p_{A_r}(x,y)$ is the position of effective aperture area of MRR in $x-y$ plane.
It can be shown that $h_{p_u}$ is proportional to the MRR aperture. This dependence suggests the use of large $A_r$ to increase the link budget. However, the switching rate of MRR modulator is inversely proportional to $A_r$. Accordingly, in practice, the active area of MRR is usually less than 1 $\textrm{cm}^2$ which is much smaller than the beamwidth $w_z$. From this point, the beam is approximately plane in the aperture of MRR and thus, Eq. \eqref{ss2} can be well approximated as
\begin{align}
\label{ss3}
h_{p_u} &= \frac{2 A_r}{\pi w_z^2} 	 
\exp\left(-\frac{2(d_{px}^2 + d_{py}^2)}{w_z^2} \right) .
\end{align}

The vector perpendicular to the aperture area is denoted by $r_z$. 
%
%The UAV tries to set the vector $r_z$ in the direction of $z$ axis by using a simple stepper motor.
Here, we assume that the UAVs orient themselves and/or use a simple servo motor (which has much lower weight and price than a stabilizer) to set the vector $r_z$ in the direction of $z$ axis.\footnote{It is assumed that the the positions of GS and UAV are known to UAV, which can be realized through periodic data exchange between them and/or by using the global navigation satellite systems.}
However, due to the inherent orientation fluctuations of UAVs, the instantaneous orientation of the vector $r_z$ is deviated from $z$ axis.
Let $\theta$ denote the instantaneous orientation deviation between vector $r_z$ and $z$ axes. As depicted in Fig. \ref{cxs}, a portion of the power collected by the MRR aperture is scattered depending on $\theta$. 
The ratio of the power reflected directly to all of the collected optical power by the MRR is denoted by $h_{MRR}$. The PDF of $h_{MRR}$ will be derived in the next section.
Then, the reflected optical power is directly back to the GS. It is further assumed that the GS uses a circular aperture for transmission and reception.
We neglect the effect of beam wandering. This assumption is valid for FSO communication with link lengths up to several kilometers \cite{huang2017free}. 
Therefore, the beam is received at the GS with a negligible deviation (due to negligible beam wandering) and thus the center of the received beam is located approximately at the center of the GS aperture. Hence, the attenuation due to geometric loss in GS aperture is 
\begin{align}
	\label{ss5}
	h_{p_g}= \frac{2 r_g^2}{w_{zg}^2}\simeq  \frac{2 r_g^2}{Z^2 \theta_\textrm{div}^2},
\end{align}
where $r_g$ is the radius of GS aperture, and $w_{zg}$ is the optical beamwidth in the GS.

In addition to the aforementioned factors, the optical power launched from the GS is also affected by atmospheric turbulence induced fading and atmospheric loss before arriving back at the GS. 
%------------------------------------
In this study, a bistatic channel is assumed where the transmitter and the receiver of GS are separated in space by more than a Fresnel zone, such that, the  round trip channel is modeled as a product of two independent turbulence channels\footnote{MRR-based FSO communication systems can be implemented in monostatic or bistatic configurations. For a monostatic system, the transmitter and receiver are colocated, whereby, there is a high correlation between the instantaneous atmospheric coefficients of the forward and backward passes, which decreases the performance, significantly. 
For instance, from the results of \cite[Fig. 4]{yang2018performance}, there is a SNR gap of more than 20 dB between a bistatic configuration with a correlation coefficient near to zero and a monostatic configuration with a correlation coefficient equal to 0.6 for a target BER of $10^{-8}$. 
%The results also show that by reducing the correlation coefficient from 0.6 to 0.2, the SNR gap decreases to 4 dB.
% In addition, beam-tracking system impairs for correlated FSO fading channels \cite{kiasaleh2006beam}. In this paper, we will show that the performance of UAV-based MRR FSO links significantly depends on the accuracy of beam-tracking system.  
On the other hand, the Fresnel zone of FSO links is in the order of one centimeter which is typically much smaller than optical half power (3 dB) beamwidth at the receiver, and thus, we can separate the transmitter and receiver of GS more than the Fresnel zone of FSO link to have a bistatic channel which causes an additional geometrical loss much less than 3 dB. However, bistatic configuration offers the advantage of preserving independent forward and backward paths \cite{andrews2005laser}. 
More importantly, beam-tracking system impairs for correlated FSO fading channels \cite{kiasaleh2006beam}. In this paper, we will show that the performance of UAV-based MRR FSO links significantly depends on the accuracy of the ground beam-tracking system. 
Therefore, bistatic is a preferable configuration in most of the work in the context of MMR-based FSO communications (for instance see \cite{li2017probability,yanmei2021modulated,nelson2015enhanced,althunibat2019performance,el2020performance,geng2014single,majumdar2015modulating,scott2004calculations}). 
In practice, the interrogator with a bistatic configuration is also adopted by Naval Research Laboratory to conduct a communication experiment over 16 km FSO links from the Chesapeake Bay to a modulated RR array \cite{5680205,plett2008free}.}\cite{andrews2005laser}.
%
%receive less power than monostatic MRR systems
%
%gathers less amount of the reflected power
%
%is shifted from the on-axis component
%
%In the bistatic systems, the forward and backward paths are separated by enough large space, and the atmospheric turbulences along the two paths are uncorrelated.
%
%When the transmitter and receiver are separated in space by more than a Fresnel zone, it is called a bistatic system.
%
%On the contrary, the bistatic configuration which demands the transmitter and receiver to be separated by the order of Fresnel zone radius, gathers less amount of the reflected power. However, bistatic configuration offers the advantage of preserving independent forward and backward paths. Therefore, the system atmospheric fading is reduced
%
%MRR FSO communication systems can be divided into two groups, monostatic and bistatic. In monostatic MRR systems both transmitter and receivers are collocated [9–11]. On the other hand, the receiver in bistatic MRR systems is shifted from the on-axis component. Despite monostatic MRR systems receive much amount of power, the fading correlation between the forward and backward paths, since they have a common axis, is a major limitation [12,13]. On the contrary, in bistatic MRR, off-axis components are detected. Therefore, as long as the transmitter and receiver are separated by the order of the Fresnel radius, independent fading can be obtained. However, bistatic MRR systems receive less power than monostatic MRR systems [14]
%------------------------------------
Therefore, the considered system model consists of two paths and thus we have two independent atmospheric turbulence induced fading effects and two independent atmospheric attenuation effects. 
%
%The instantaneous atmospheric turbulence coefficients of BS-to-UAV and UAV-to-BS are respectively denoted by $h_{a_{bu}}$ and $h_{a_{ub}}$ and can be formulated as
%
Let $h_{a_{gu}}$ and $h_{a_{ug}}$ denote the instantaneous atmospheric turbulence coefficients of GS-to-UAV and UAV-to-GS, respectively. Further, let $h_{l_{gu}}$ and $h_{l_{ug}}$ denote the atmospheric attenuation of GS-to-UAV and UAV-to-GS, respectively.
The atmospheric attenuation is typically modeled by the Beer-Lambert law as \cite{ghassemlooy2019optical} 
\begin{align}
	\label{ss6}
	h_{l_{gu}}=h_{l_{ug}}=\exp(-Z\zeta),
\end{align}
where $\zeta$ is the scattering coefficient and is a function of visibility. Notice that $h_{l_{gu}}$ and $h_{l_{ug}}$ are equal since the parameters $\zeta$ and $Z$ are the same for both links.
The log-normal (LN) and Gamma-Gamma (GG) distributions are good candidates to efficiently model weak to moderate and moderate to strong ranges of atmospheric turbulence conditions \cite{ghassemlooy2019optical,andrews2005laser}
Under weak to moderate turbulence conditions, we use LN distribution to model $h_{a_{gu}}$ and $h_{a_{ug}}$ as \cite{andrews2005laser}
\begin{align}
	\label{ss9}
	f_L\left(h_{a_{n}}\right) = \frac{1}{2h_{a_{n}}\sqrt{2\pi \sigma_L^2}  }
		\exp \left(- \frac{\left( \ln(h_{a_{n}}) + 2\sigma_L^2 \right)^2} {8\sigma_L^2} \right),
\end{align}
where $n\in\{ub,bu\}$, $\sigma_L^2\simeq \sigma_R^2/4$ is the SD of the log-normal distribution and $\sigma_R^2$ is the  Rytov variance which can be obtained for two nodes with different heights as \cite[p. 509]{andrews2005laser}
%\begin{align}
%	\label{sss1}
%	\sigma_L^2 = 0.30545 k^{7/6} C_n^2(Z_h) Z^{11/6},
%\end{align} 
\begin{align}
	\label{sss1}
	&\sigma_R^2 = 9 \left( {2\pi}/{\lambda} \right)^{7/6}  \left( Z/Z_{h_d} \right)^{11/6} \\
	&~~~\times \int_{Z_{h_g}}^{Z_{h_u}} 
	C_n^2(Z_h) \left( 1 - \frac{Z_h-Z_{h_g}}{Z_{h_d}}  \right)^{5/6}  (Z_h-Z_{h_g})^{5/6}  \textrm{d}Z_h. \nonumber
\end{align}
In \eqref{sss1}, $Z_{h_u}$ and $Z_{h_g}$ denote the height of UAV and GS, respectively, and $Z_{h_d}=Z_{h_u}-Z_{h_g}$ denote the difference height between the UAV and GS, and
\begin{align}
	&C_n^2(Z_h) = 0.00594 (\mathcal{V}/27)^2 \left(10^{-5} Z_h \right)^{10}
	\exp\left( -\frac{Z_h}{1000} \right) \nonumber \\
	&~~~ + 2.7 \times 10^{-16}   \exp\left( -\frac{Z_h}{1500} \right) 
	+ C_n^2(0) \exp\left( -\frac{Z_h}{100} \right), \nonumber
\end{align}
is the refractive-index structure parameter at height $Z_h$ which characterizes the atmospheric turbulence, $\mathcal{V}$ (in m/s) is the speed of strong wind and $C_n^2(0)$ (in $\textrm{m}^{-2/3}$) is a strong nominal ground turbulence level.
Under moderate to strong turbulence conditions, we use GG distribution to model the random variables (RVs) $h_{a_{gu}}$ and $h_{a_{ug}}$ as \cite{andrews2005laser}
\begin{align}
	\label{ss8}
	f_G\left(h_{a_{n}}\right) = \frac{2(\alpha\beta)^{\frac{\alpha+\beta}{2}}}{\Gamma(\alpha)\Gamma(\beta)}
	h_{a_n}^{^{\frac{\alpha+\beta}{2}-1}}
	k_{\alpha-\beta}  \left( 2\sqrt{\alpha\beta h_{a_{n}}}\right),
\end{align}
where $\Gamma(\cdot)$ is the Gamma function and $k_m(\cdot)$ is the modified Bessel function of the second kind of order $m$. Also, $\alpha$ and $\beta$ are respectively the effective number of large-scale and small-scale eddies, which depend on Rytov variance $\sigma_R^2$ \cite{andrews2005laser}.

From the aforementioned results and as graphically depicted in Fig. \ref{r3}, the instantaneous GS-to-UAV-to-GS channel coefficient is formulated as
\begin{align}
\label{s}
h = h_{l_{ug}} h_{l_{gu}}  h_{a_{ug}}  h_{a_{gu}}  h_{p_u}  h_{p_g}  h_{MRR}.
\end{align}
Finally, the instantaneous end-to-end SNR is obtained as \cite{ghassemlooy2019optical}
\begin{align}
	\label{p1}
	\Upsilon = \frac{2 R^2 P_t^2 h^2}{\sigma_n^2},
\end{align}
where $P_t$ denotes the transmitted optical power, $R$ denotes the photo-detector responsivity, and $\sigma_n^2$ is the variance of additive thermal noise. 
%\newpage
%
%%%%%%%%%%%%%%%%%%%%%%%%%%%%%%%%%%%%%%%%%%%%%%%%%%%%%%%%%%%%%%%%
%%%%%%%%%%%%%%%%%%%%%%%%%%%%%%%%%%%%%%%%%%%%%%%%%%%%%%%%%%%%%%%%
%%%%%%%%%%%%%%%%%%%%%%%%%%%%%%%%%%%%%%%%%%%%%%%%%%%%%%%%%%%%%%%%
%%%%%%%%%%%%%%%%%%%%%%%%%%%%%%%%%%%%%%%%%%%%%%%%%%%%%%%%%%%%%%%%
%channel_coefficients

%------------------------------------------
%------------------------------------------
\section{Channel Modeling}
%------------------------------------------
%------------------------------------------
%Next, the analytical channel model of the considered communication system is derived for both weak to moderate and moderate to strong conditions. To obtain the channel model, we first need to find $h_{MRR}$.
%

\begin{figure*}
	\centering
	\subfloat[] {\includegraphics[width=2.1 in]{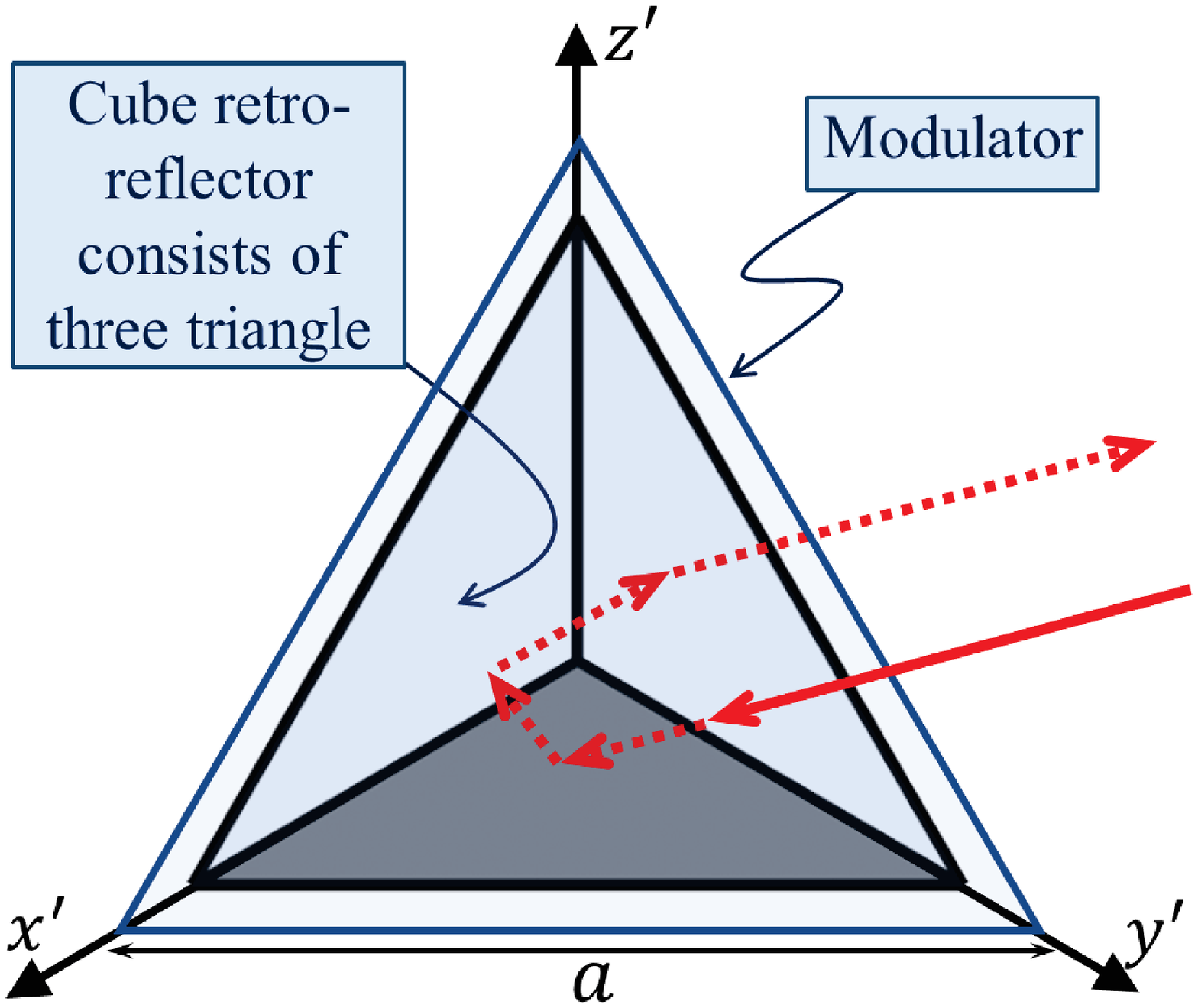}
		\label{cx1}
	}
	\hfill
	\subfloat[] {\includegraphics[width=2.9 in]{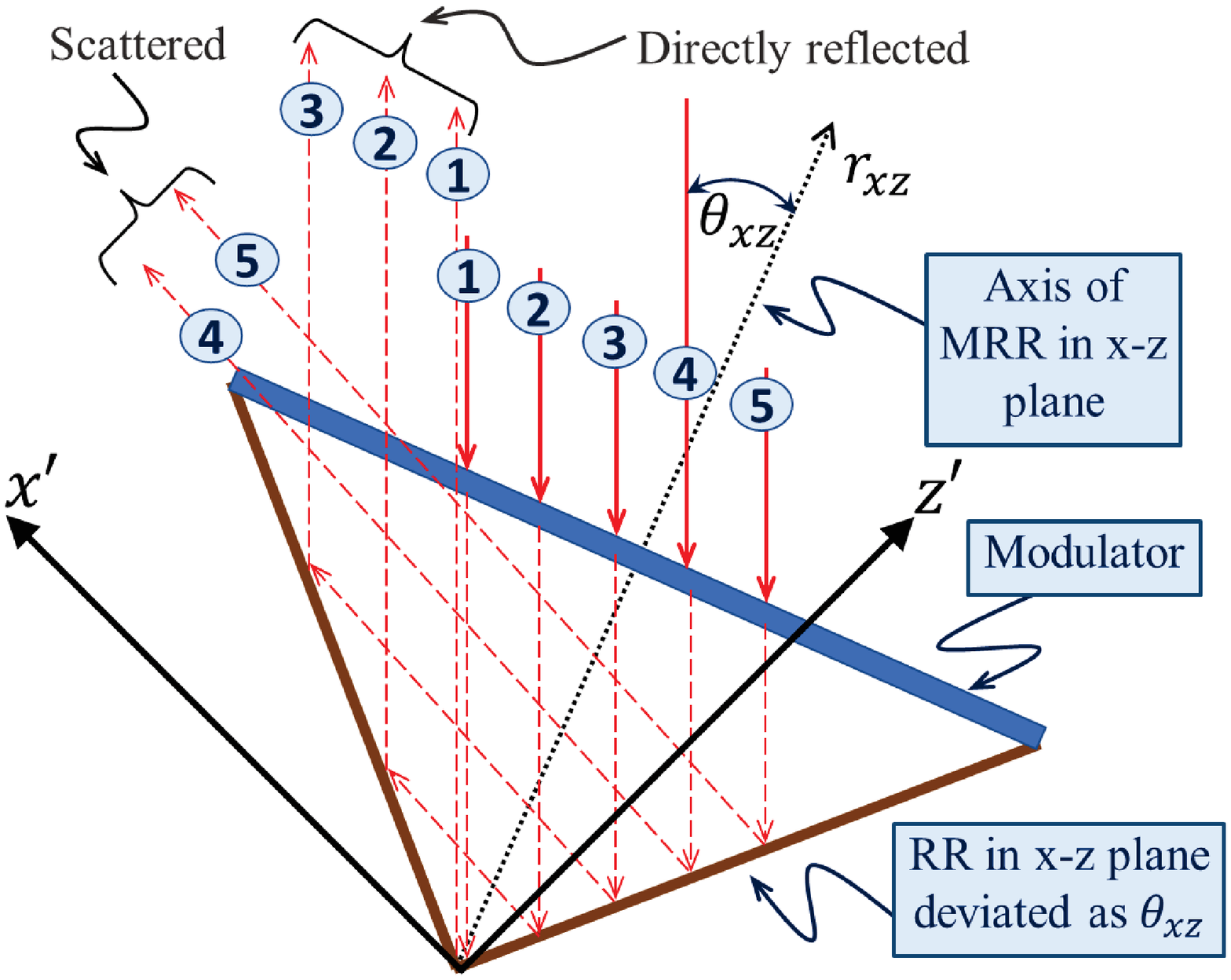}
		\label{cx2}
	}
	\hfill
	\subfloat[] {\includegraphics[width=1.8 in]{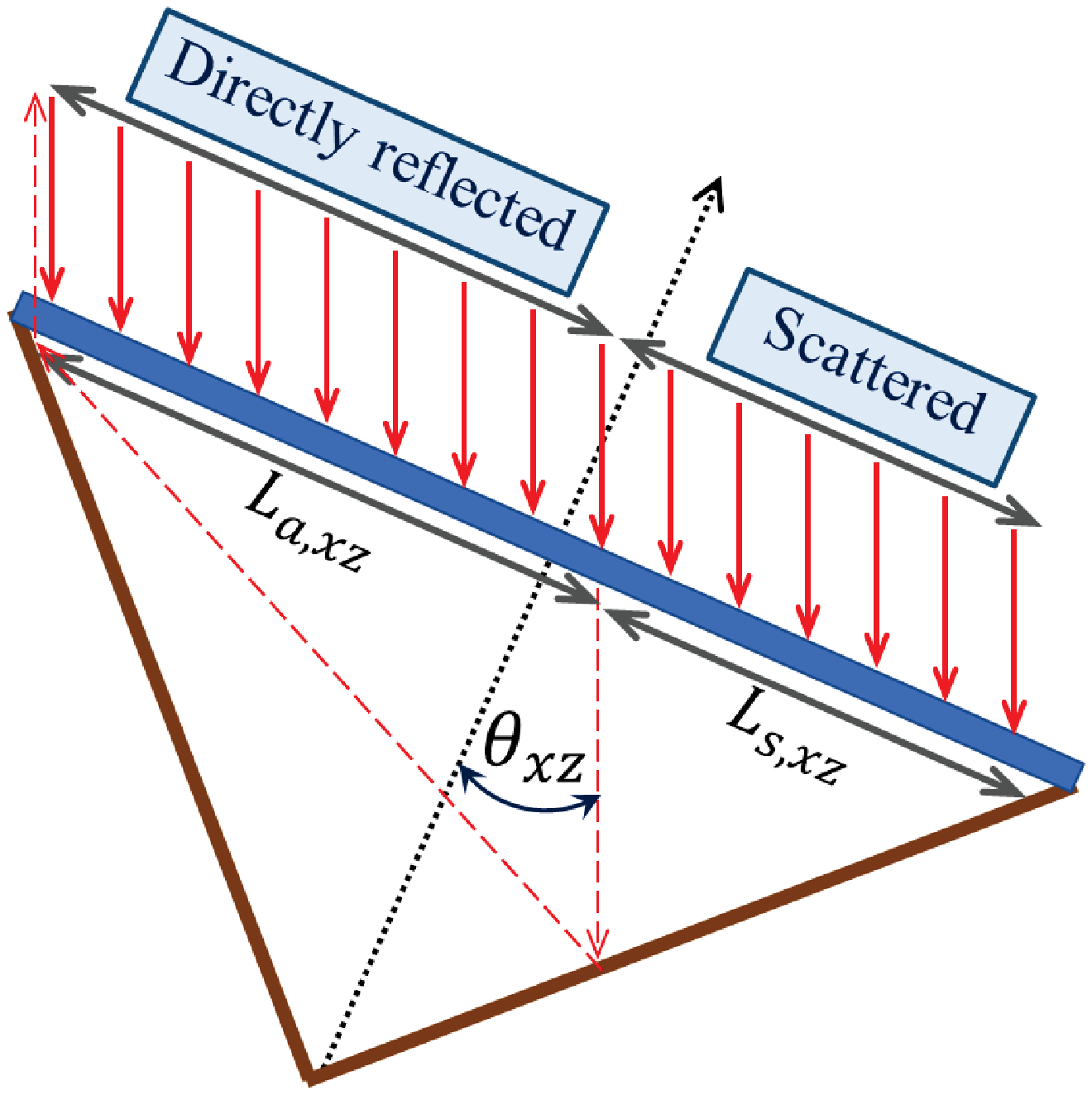}
		\label{cx3}
	}
	\caption{3D illustration of  MRR configuration:
		(a) the MRR consists of three perpendicular triangular mirrors that are located in Coordinate planes $x'-y'$, $x'-z'$, and $y'-z'$; 
		(b) the angles between the incident laser beam and the vectors perpendicular to the aperture area in $x'-z'$ plane;
		(c) A fraction of collected laser power by aperture is directly reflected to the GS which is a function of RVs $\theta_{xy}$, $\theta_{xz}$, and $\theta_{yz}$.
	%3D illustration of antenna pattern generated by a uniform Nq × Nq antenna array: (a) showing a Nq × Nq square array antenna arranged in [x; y] plane; (b) 3D actual antenna pattern generated by a square array antenna arranged on [x; y] plane; (c) approximated antenna pattern obtained by (13)
     }
	\label{cxs}
\end{figure*}

\begin{figure}
	\begin{center}
		\includegraphics[width=3.2 in]{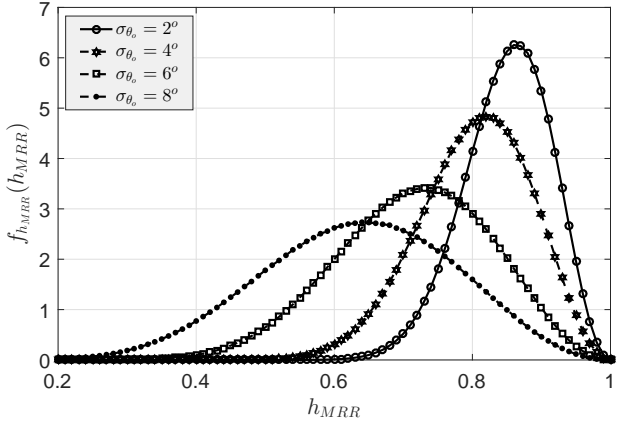}
		\caption{The distribution of $h_{MRR}$ obtained using simulation for different values of $\sigma_{\theta_o}$.  }
		\label{cx4}
	\end{center}
\end{figure}

The considered MRR consists of three perpendicular triangular mirrors that are located in Coordinate planes $x'-y'$, $x'-z'$, and $y'-z'$, as depicted in Fig. \ref{cx1} where $[x'; y'; z']$ is a new Cartesian coordinate system.
Let the vectors $r_{xy}$, $r_{xz}$, and $r_{yz}$ denote the vectors perpendicular to the aperture area in $x'-y'$, $x'-z'$, and $y'-z'$, respectively. 
Also, the angles between the incident laser beam and the vectors $r_{xy}$, $r_{xz}$, and $r_{yz}$ in  $x'-y'$, $x'-z'$, and $y'-z'$ are denoted respectively by $\theta_{xy}\sim\mathcal{N}\left(0,\sigma_{\theta_{o}}^2\right)$, 
   $\theta_{xz}\sim\mathcal{N}\left(0,\sigma_{\theta_{o}}^2\right)$, and 
   $\theta_{yz}\sim\mathcal{N}\left(0,\sigma_{\theta_{o}}^2\right)$ as depicted in Fig. \ref{cx2}.
A fraction of the collected laser power by aperture is directly reflected to the GS which is a function of RVs $\theta_{xy}$, $\theta_{xz}$, and $\theta_{yz}$.
Due to the independence of RVs $\theta_{xy}$, $\theta_{xz}$, and $\theta_{yz}$, we have
\begin{align}
	\label{g1}
	h_{MRR}= h_{MRR_{xy}}(\theta_{xy})\, h_{MRR_{xz}}(\theta_{xz})\, h_{MRR_{yz}}(\theta_{yz}),
\end{align}
where $h_{MRR_{xy}}$, $h_{MRR_{xz}}$, and $h_{MRR_{yz}}$ are the ratio of direct reflected optical power to all of the collected optical power by MRR in planes $x'-y'$, $x'-z'$, and $y'-z'$, respectively.
For a better understanding, and for a given $\theta_{xz}$ in $x'-z'$ plane, we have graphically shown in Figs. \ref{cx2} and \ref{cx3} which category of input optical signal is directly reflected (denoted by $L_{a,xz}$) and which category is scattered (denoted by $L_{s,xz}$). From this, we have
\begin{align}
	\label{g2}
	h_{MRR_{n}}(\theta_n) = \frac{ L_{a,n} }{ L_{a,n} + L_{s,n}}  = 1-\tan(\theta_n).
\end{align}
for $n\in\{xy,xz,yz\}$. From \eqref{g1} and \eqref{g2}, finding a closed-form expression for the PDF of $h_{MRR}$ is very difficult if not impossible. However, for system analysis, we need to know the PDF of $h_{MRR}$ denoted by $f_{h_{MRR}}(h_{MRR})$.
Accordingly, in the sequel, $f_{h_{MRR}}(h_{MRR})$ is approximated by two simple models based on the mean and variance of $h_{MRR}$. 
In Fig. \ref{cx4}, the distribution of $h_{MRR}$ is obtained by using simulations for different values of $\sigma_{\theta_o}$.
As shown in Fig. \ref{cx4}, the mean and variance of $h_{MRR}$ (denoted by $\mu_{MRR}$ and $\sigma_{MRR}^2$, respectively) are functions of the angular instability of the UAV characterized by $\sigma_{\theta_{o}}$. 
In Table \ref{tab1}, $\mu_{MRR}$ and $\sigma_{MRR}^2$ of $h_{MRR}$ are obtained by using simulations for different values of $\sigma_{\theta_o}$.
For the rest of $\sigma_{\theta_o}$ values, the corresponding mean and variance values can be obtained by interpolation from the given values. For instance, from the results of Table II, for $\sigma_{t_o}=1^o$ and $\sigma_{t_o}=3^o$ we have $\mu_{MRR}=0.95$ and $\mu_{MRR}=0.85$, respectively. Now, for $\sigma_{t_o}=2^o$, by using interpolation, we obtain $\mu_{MRR}\simeq \frac{0.95+0.85}{2}=0.9$.

%
%%%%%%%%%%%%%%%%%%%%%%%%%%%%%%%%%%%%%%%%%%%%%%%%%%%%%%%%%%%%%%%%
%%%%%%%%%%%%%%%%%%%%%%%%%%%%%%%%%%%%%%%%%%%%%%%%%%%%%%%%%%%%%%%% VERSUS W_Z
%%%%%%%%%%%%%%%%%%%%%%%%%%%%%%%%%%%%%%%%%%%%%%%%%%%%%%%%%%%%%%%%
%%%%%%%%%%%%%%%%%%%%%%%%%%%%%%%%%%%%%%%%%%%%%%%%%%%%%%%%%%%%%%%%  
%

%
%%%%%%%%%%%%%%%%%%%%%%%%%%%%%%%%%%%%%%%%%%%%%%%%%%%%%%%%%%%%%%%%
%%%%%%%%%%%%%%%%%%%%%%%%%%%%%%%%%%%%%%%%%%%%%%%%%%%%%%%%%%%%%%%%
%%%%%%%%%%%%%%%%%%%%%%%%%%%%%%%%%%%%%%%%%%%%%%%%%%%%%%%%%%%%%%%%
%%%%%%%%%%%%%%%%%%%%%%%%%%%%%%%%%%%%%%%%%%%%%%%%%%%%%%%%%%%%%%%%
%

%
%
%%%%%%%%%%%%%%%%%%%%%%%%% TABEL of simple UAV system
\begin{table*}
	\def\tablename{Table}
	\centering
	\caption{Mean and SD values of RV $h_{MRR}$ obtained using simulation for different values of $\sigma_{\theta_o}$.
		%$\sigma_{\rm to}~\textrm{and}~\sigma_{\rm ro}$	when $P_t=15~{\rm dBm}$.
	}
	% $\theta'_\textrm{tx}=\theta'_\textrm{ty}=\theta'_\textrm{rx}=\theta'_\textrm{ry}=1^o$
	%\resizebox{\textwidth}{!}{
	\begin{tabular}{|c||   
			c|c|c|c|c|c|c|c|c|c|c|}
		\cline{1-12}
		%%------------------------------------------------------------- 
		%%-------------------------------------------------------------
		%%------------------------------------------------------------- Title
		$\sigma_{\theta_o}$ & $1^o$ & $2^o$ & $3^o$ & $4^o$ & $5^o$ & $6^o$ & $7^o$ & $8^o$ & $9^o$ & $10^o$ & $11^o$ \\
		\hline \hline 
		%%------------------------------------------------------------- 
		%%-------------------------------------------------------------
		%%------------------------------------------------------------- Title
		%%------------------------------------------------------------- Title
		$\mu_{MRR}$&
		0.96 & 0.93 & 0.89 & 0.86 & 0.83 & 0.8 & 0.76 & 0.73 & 0.70 & 0.66 & 0.62 \\ \hline  
		%%------------------------------------------------------------- 
		%%-------------------------------------------------------------
		%%------------------------------------------------------------- Title
		%%------------------------------------------------------------- Title
		% .0178   .035      0.052   0.066   .083   .094  .11  .12   .13    .145  .158
		$\sigma_{MRR}$&
		0.0178  & .035 &     0.052  & 0.066 &  0.083 &  0.094 & 0.11 & 0.12  & 0.13  &  0.145 & 0.158\\
		\hline
	\end{tabular}%}
	\label{tab1}%
\end{table*}%
%%%%%%%%%%%%%%%%%%%%%%%%%%%%%%%%%%%%
%%%%%%%%%%%%%%%%%%%%%%%%%%%%%%%%%%%
%
%

%------------------------------------------
%------------------------------------------
\subsection{Weak Turbulence Conditions}
%------------------------------------------
%------------------------------------------
We use the log-normal distribution to model atmospheric turbulence induced fading that is valid for weak to moderate turbulence conditions.

%
%------------- Proposition 1 ----------------------------------------------------------------
%------------- Proposition 1 ----------------------------------------------------------------
%------------- Proposition 1 ----------------------------------------------------------------
%------------- Proposition 1 ----------------------------------------------------------------
%------------- Proposition 1 ----------------------------------------------------------------
{\bf Theorem 1.}  {\it  The distribution of $h$ under weak to moderate atmospheric turbulence conditions is derived as}
\begin{align}
	\label{sr8}
	& f_h(h) =   C_4  h^{K-1} Q\left(  \frac{\ln(h) + C_5}{\sqrt{C_1}}   \right), 
\end{align}
{\it where}
\begin{align}
	%\label{sr9}
	\left\{
	\begin{array}{ll}
		&\!\!\!\!\!\!\!C_1 = \ln\left(1+\frac{\sigma_{MRR}^2}{\mu_{MRR}^2}\right) + 8\sigma_L^2, \\
		&\!\!\!\!\!\!\!C_2 = \ln\left(\frac{\sqrt{\mu_{MRR}^2+\sigma_{MRR}^2}}{\mu_{MRR}^2} \right)+4\sigma_L^2, \\
		&\!\!\!\!\!\!\!C_3 = \frac{\pi w_z^2}{2A_r h_c}, \\
		&\!\!\!\!\!\!\!C_4 = K C_3^K   \exp\left(  \frac{C_1 K^2 - 2K C_2  }{2}   \right),\\
		&\!\!\!\!\!\!\!C_5  = \ln(C_3) + C_1K + C_2, \\
		&\!\!\!\!\!\!\!K=\frac{w_z^2}{Z^2 \sigma_{\theta_e}^2},~~~\& ~~~ h_c = h_{l_{ug}} h_{l_{gu}} h_{p_g},
	\end{array}
	\right. \nonumber 
\end{align}
{\it and $Q(\cdot)$ is the well-known  Q-function.}

%%%%%%%%%%%%%%%%%%%%%%%%%%%%%%%%%%
%%%%%%%%%%%%%%%%%%%%%%%%%%% begin PROOF %%%%%%%%%%%%%%%%%%%%%%%%%%%%%%%%%%%%%%%%%%%%%%%%%%%%%%%%%%%%%%%%%%%%%%%%%%%%%%%%%%%%%%%%%%%%%%%%%%%%%%%%%%%%%%%%%%%%%%%%%%
\begin{IEEEproof}
	Please refer to Appendix \ref{weak_dist}.
\end{IEEEproof}

As we will see in the next section, the accuracy of the derived analytical channel model will be validated by employing Monte Carlo simulations.
Moreover, the channel model provided in \eqref{sr8} is very simple and tractable which allows us to easily analyze the performance of the considered MRR-based FSO system without performing time consuming simulations. Next, the BER and outage probability of the considered system under weak to moderate atmospheric turbulence conditions are derived.

%
%------------- Proposition 1 ----------------------------------------------------------------
%------------- Proposition 1 ----------------------------------------------------------------
%------------- Proposition 1 ----------------------------------------------------------------
%------------- Proposition 1 ----------------------------------------------------------------
%------------- Proposition 1 ----------------------------------------------------------------
{\bf Proposition 1.}  {\it  The {CDF} of $h$ under weak to moderate atmospheric turbulence conditions is derived as}
\begin{align}
	\label{sf1}
	F_h(h) &= \frac{C_4}{K} e^{-KC_5}  \left[ 
	e^{K\ln(h) + KC_5}    Q\left( \frac{\ln(h)+C_5}{\sqrt{C_1}} \right)  \right.\nonumber \\
	%---
	&~~~\left.+~ e^{K^2C_1/2}  Q\left( \frac{K C_1-\ln(h)-C_5}{\sqrt{C_1}} \right)
	\right].
\end{align}
\begin{IEEEproof}
The CDF of $h$ is defined as 
\begin{align}
	\label{cdf_general}
	F_h(h) = \int_0^h f_h(x)dx.
\end{align} 
Using \eqref{sr8}, we have
\begin{align}
\label{df1}
F_h(h) =   C_4  \int_0^h x^{K-1} Q\left(  \frac{\ln(x) + C_5}{\sqrt{C_1}}   \right) dx.
\end{align}	
Applying a change of variable $y=\ln(x) + C_5$ and using \eqref{df1} and \cite[eq. (06.27.21.0011.01)]{wolfram}, the closed-form expression for CDF of $h$ is derived in \eqref{sf1}.
\end{IEEEproof}
%
%
%

%------------- Proposition 2 ----------------------------------------------------------------
%------------- Proposition 2 ----------------------------------------------------------------
%------------- Proposition 2 ----------------------------------------------------------------
%------------- Proposition 2 ----------------------------------------------------------------
%------------- Proposition 2 ----------------------------------------------------------------
{\bf Proposition 2.}  {\it  The PDF and CDF of end-to-end SNR under weak to moderate atmospheric turbulence conditions are derived as}
\begin{align}
	\label{sf2}
	f_\Upsilon(\Upsilon) = \frac{C_4}{2 \Upsilon_1^{K/2}} 
	\Upsilon^{K/2-1}    Q\left(\! \frac{\ln(\Upsilon) - \ln(\Upsilon_1) \!+\! 2C_5}{2\sqrt{C_1}} \!  \right),
\end{align}
{\it and}
\begin{align}
	\label{sf3}  % \left(\frac{\Upsilon}{\Upsilon_1} \right)
	F_\Upsilon(\Upsilon) &= \frac{C_4}{K} e^{-KC_5}  \left[ 
	Q\left( \frac{\ln(\Upsilon) - \ln(\Upsilon_1)+2C_5}{2\sqrt{C_1}} \right)
	 \right.  \\
	%-----------  
	&~\times \exp\left(  \frac{K\ln(\Upsilon) - K\ln(\Upsilon_1) + 2KC_5}{2} \right)  \nonumber \\
	%---
	&~\left.+~ e^{K^2C_1/2}  Q\left( \frac{2K C_1-\ln(\Upsilon)-\ln(\Upsilon_1)-2C_5}{2\sqrt{C_1}} \right)
	\right], \nonumber
\end{align}
{\it where $\Upsilon_1 = \frac{2 R^2 P_t^2 }{\sigma_n^2}$}.
\begin{IEEEproof}
	Based on \eqref{p1}, we have
	\begin{align}
		\label{k1}
		F_\Upsilon(\Upsilon) = \textrm{Prob}\!\left\{\!\frac{2 R^2 P_t^2 }{\sigma_n^2} h^2<\Upsilon\!\right\} 
		= \textrm{Prob}\!\left\{\!\Upsilon_1 h^2<\!\Upsilon\!\right\}.
	\end{align}
	Using \eqref{sf1} and \eqref{k1}, the CDF of $\Upsilon$ is derived in \eqref{sf3}. Now, by differentiating \eqref{sf3} with respect to $\Upsilon$ and after some manipulations, the closed-form analytical expression for the PDF of $\Upsilon$ is derived in \eqref{sf2}.
\end{IEEEproof}

In slow fading channels, which is a valid assumption for channel condition in FSO links \cite{khalighi2014survey}, the outage probability (i.e., the probability that the instantaneous end-to-end SNR falls below a threshold $\Upsilon_\textrm{th}$) is the most appropriate performance metric. It is formulated as $\mathbb{P}_\textrm{out} = F_\Upsilon(\Upsilon_\textrm{th})$. Accordingly, the results of Proposition 2 let us to easily compute outage probability by substituting $\Upsilon_\textrm{th}$ instead of $\Upsilon$ in \eqref{sf3}.

%
%------------- Proposition 1 ----------------------------------------------------------------
%------------- Proposition 1 ----------------------------------------------------------------
%------------- Proposition 1 ----------------------------------------------------------------
%------------- Proposition 1 ----------------------------------------------------------------
%------------- Proposition 1 ----------------------------------------------------------------
{\bf Proposition 3.}  {\it The closed-form expression for BER of the considered MRR-based FSO communication for On–off keying (OOK) modulation over weak to moderate turbulence conditions is derived as}
\begin{align}
	\label{a4}
	& \mathbb{P}_e \simeq  L_1 L_2 \Bigg[
	\frac{1}{b} \left( e^{\frac{b^2}{2 a^2}} \textrm{erfc}\left( \frac{b}{2a}-aL_3 \right)   +  e^{b L_3} \textrm{erfc}(aL_3)   \right) \nonumber \\
	%--------------------
	&-\sum_{m=0}^M    \frac{L_{1m}}{b_m} \left( e^{\frac{b_m^2}{2 a^2}} \textrm{erfc}\left( \frac{b_m}{2a}-aL_3 \right)   +  e^{b_m L_3} \textrm{erfc}(aL_3)   \right)
	\Bigg], 	
\end{align}

\begin{align}
	%\label{sr9}
	\left\{
	\begin{array}{ll}
		&\!\!\!\!\!\!\! L_1 = \frac{C_4}{4 \Upsilon_1^{K/2}}, \\
		%-------
		&\!\!\!\!\!\!\! L_2 = \frac{1}{2} \exp\left(  \frac{K(\ln(\Upsilon_1) - 2C_5)}{2}  \right), \\
		&\!\!\!\!\!\!\! L_3 = \ln(\Upsilon_\textrm{max}) -\ln(\Upsilon_1) + 2C_5, \\
		&\!\!\!\!\!\!\! L_{1m} = \frac{(-1)^m }   {m!\sqrt{\pi}(2m+1)2^{m-\frac12}  }    \exp\left(  \frac{(2m+1)(\ln(\Upsilon_1) - 2C_5)}{2} \right), \\
		&\!\!\!\!\!\!\! a = \frac{1}{2\sqrt{2C_1}},~~~\&~~~b=\frac{K}{2},~~~\&~~~ b_m =\frac{K+2m+1}{2}.
	\end{array}
	\right. \nonumber 
\end{align}

%%%%%%%%%%%%%%%%%%%%%%%%%%%%%%%%%%
%%%%%%%%%%%%%%%%%%%%%%%%%%% begin PROOF %%%%%%%%%%%%%%%%%%%%%%%%%%%%%%%%%%%%%%%%%%%%%%%%%%%%%%%%%%%%%%%%%%%%%%%%%%%%%%%%%%%%%%%%%%%%%%%%%%%%%%%%%%%%%%%%%%%%%%%%%%
\begin{IEEEproof}
	Please refer to Appendix \ref{weak_BER}.
\end{IEEEproof}

The accuracy of the approximated BER provided in \eqref{a4} increases by increasing $M$ and $\Upsilon_\textrm{max}$. Notice that the term of $e^{\frac{b_m^2}{2 a^2}}$ in \eqref{a4} becomes uncountably large by increasing $M$. for large values of $M$ and $\Upsilon_\textrm{max}$. Therefore, due to the computational limitations, 
we will limit our simulations to moderate values of $M$ and $\Upsilon_\textrm{max}$. After an exhaustive search, we found that the BER is obtained with good accuracy for $M=20$ and $\Upsilon_\textrm{max}=4$.

%------------------------------------------moderation to strong turbulence conditions
%------------------------------------------
\subsection{Moderate to Strong Turbulence Conditions}
%------------------------------------------
%------------------------------------------
We use the GG distribution to model atmospheric turbulence induced fading that is valid for moderate to strong turbulence conditions.

%
%------------- Proposition 1 ----------------------------------------------------------------
%------------- Proposition 1 ----------------------------------------------------------------
%------------- Proposition 1 ----------------------------------------------------------------
%------------- Proposition 1 ----------------------------------------------------------------
%------------- Proposition 1 ----------------------------------------------------------------
{\bf Theorem 2.}  {\it  The PDF and CDF of $h$ under moderate to strong atmospheric turbulence conditions are derived as}
\begin{align}
	\label{st8}
	&f_{h}(h) = B_s  B_n  \sum_{n=1}^N\\
	%
	%-------------
	&\left[
	G^{6,0}_{2,6}\left( B_n' h \bigg|
	\begin{array}{c}
		K,1
		\\
		0,\alpha-1,\beta-1,K-1,\alpha-1,\beta-1 
	\end{array}
	\right)  \right. \nonumber \\
	%--------------
	&-\left.G^{6,0}_{2,6}\left(B_n'' h \bigg|
	\begin{array}{c}
		K,1
		\\
		0,\alpha-1,\beta-1,K-1,\alpha-1,\beta-1 
	\end{array}
	\right) \right], \nonumber
\end{align}
and
\begin{align}
	\label{cdf_strong}
	&F_{h}(h) = B_s  B_n h \sum_{n=1}^N\\
	%
	%-------------
	&\left[
	G^{6,1}_{3,7}\left( B_n' h \bigg|
	\begin{array}{c}
		0,K,1
		\\
		0,\alpha-1,\beta-1,K-1,\alpha-1,\beta-1,-1 
	\end{array}
	\right)  \right. \nonumber \\
	%--------------
	&- \left.G^{6,1}_{3,7}\left(B_n'' h \bigg|
	\begin{array}{c}
		0,K,1
		\\
		0,\alpha\!-\!1,\beta\!-\!1,K-1,\alpha-1,\beta-1,-1 
	\end{array}
	\right)  \right], \nonumber 
\end{align}
where
$$ V_n= \left\{
\begin{array}{ll}
	&\!\!\!\!\!\!\!2 \mu_{MRR}-1, ~~~~~~~~~~~~~~~~~~~ n=1,\\
	%-------
	&\!\!\!\!\!\!\! V_{n-1}+\frac{(2-2\mu_{MRR})n}{N}, ~~~ n\in\{2,...,N+1\},
\end{array}
\right.$$
and $U(x) = \left\{
\begin{array}{ll}
	&\!\!\!\!\!\!\!1,~~~x>0, \\
	%-------
	&\!\!\!\!\!\!\!0,~~~x<0,
\end{array}
\right. $ is the Heaviside step function,
$B_n'=\frac{ \pi w_z^2\alpha^2 \beta^2 } {2A_r h_c V_{n+1}}$, 
$B_n''=\frac{ \pi w_z^2\alpha^2 \beta^2 } {2A_r h_c V_{n}}$, and
$B_s=\frac{K  \left(\pi w_z^2 \alpha^2 \beta^2  \right)}
{(\Gamma(\alpha)\Gamma(\beta))^2  \left(2A_r h_c  \right)}$.

%%%%%%%%%%%%%%%%%%%%%%%%%%%%%%%%%%
%%%%%%%%%%%%%%%%%%%%%%%%%%% begin PROOF %%%%%%%%%%%%%%%%%%%%%%%%%%%%%%%%%%%%%%%%%%%%%%%%%%%%%%%%%%%%%%%%%%%%%%%%%%%%%%%%%%%%%%%%%%%%%%%%%%%%%%%%%%%%%%%%%%%%%%%%%%
\begin{IEEEproof}
	Please refer to Appendix \ref{strong_dist}.
\end{IEEEproof}

The coefficients $B_n$ in \eqref{st8} depend on the parameters $\sigma_{\theta_o}$ and the number of sectors denoted by $N$.
Obviously, the accuracy of the proposed model in \eqref{st8} directly depends on the number of sectors $N$, and for sufficiently large values of $N$, an exact match between simulations and analysis can be achieved at the cost of higher complexity. Hence, choosing an optimal
value for $N$ involves a trade-off between  complexity and accuracy. 
We will show, via simulations, that $N=8$ achieves sufficient accuracy.  
In Table \ref{sec_tab}, the coefficients $B_n$ obtained using simulation for different values of $\sigma_{\theta_o}$ and $N=8$.
For the rest of $\sigma_{\theta_o}$, the corresponding coefficients $B_n$ can be obtained by interpolation.%, are presented for the given values.

%
%%%%%%%%%%%%%%%%%%%%%%%%% TABEL of simple UAV system
\begin{table}
	\def\tablename{Table}
	\centering
	\caption{Coefficient $B_n$ for $N=8$ obtained using simulation for different values of $\sigma_{\theta_o}$.
		%$\sigma_{\rm to}~\textrm{and}~\sigma_{\rm ro}$	when $P_t=15~{\rm dBm}$.
	}
	% $\theta'_\textrm{tx}=\theta'_\textrm{ty}=\theta'_\textrm{rx}=\theta'_\textrm{ry}=1^o$
	%\resizebox{\textwidth}{!}{
	\begin{tabular}{|c||   
			c|c|c|c|c|c|}
		\cline{1-7}
		%%------------------------------------------------------------- 
		%%-------------------------------------------------------------
		%%------------------------------------------------------------- Title
		$\sigma_{\theta_o}$&$1^o$&
		$3^o$&$5^o$&$7^o$& $9^o$&
		$11^o$ \\
		\hline \hline 
		%%------------------------------------------------------------- 
		%%-------------------------------------------------------------
		%%------------------------------------------------------------- Title
		%%------------------------------------------------------------- Title
		$B_1$&
		2.63 & 0.85 & 0.47 & 0.29 & 0.19 & 0.1 \\ \hline  
		%%------------------------------------------------------------- 
		%%-------------------------------------------------------------
		%%------------------------------------------------------------- Title
		%%------------------------------------------------------------- Title
		$B_2$&
		5.74 & 1.99 & 1.24 & 0.91 & 0.72 & 0.58 \\
		\hline
		%%------------------------------------------------------------- 
		%%-------------------------------------------------------------
		%%------------------------------------------------------------- Title
		%%------------------------------------------------------------- Title
		$B_3$&
		10.37 & 3.73 & 2.42 & 1.87 & 1.58 & 1.42 \\
		\hline
		%%------------------------------------------------------------- 
		%%-------------------------------------------------------------
		%%------------------------------------------------------------- Title
		%%------------------------------------------------------------- Title
		$B_4$&
		15.2 & 5.49 & 3.56 & 2.75 & 2.3 & 2.07 \\
		\hline
		%%------------------------------------------------------------- 
		%%-------------------------------------------------------------
		%%------------------------------------------------------------- Title
		%%------------------------------------------------------------- Title
		$B_5$&
		17.8 & 6.19 & 3.9 & 2.94 & 2.4 & 2.06 \\
		\hline
		%%------------------------------------------------------------- 
		%%-------------------------------------------------------------
		%%------------------------------------------------------------- Title
		%%------------------------------------------------------------- Title
		$B_6$&
		14.7 & 4.99 & 3.03& 2.2 & 1.73 & 1.42 \\
		\hline
		%%------------------------------------------------------------- 
		%%-------------------------------------------------------------
		%%------------------------------------------------------------- Title
		%%------------------------------------------------------------- Title
		$B_7$&
		7.05 & 2.45 & 1.44& 1 & 0.76& 0.6 \\
		\hline
		%%------------------------------------------------------------- 
		%%-------------------------------------------------------------
		%%------------------------------------------------------------- Title
		%%------------------------------------------------------------- Title
		$B_8$&
		1.26 & 0.4 & 0.23& 0.15 & 0.11& 0.08 \\
		\hline
	\end{tabular}%}
	%}
	\label{sec_tab}%
\end{table}%
%%%%%%%%%%%%%%%%%%%%%%%%%%%%%%%%%%%%
%%%%%%%%%%%%%%%%%%%%%%%%%%%%%%%%%%%
%
%
%------------- Proposition 2 ----------------------------------------------------------------
%------------- Proposition 2 ----------------------------------------------------------------
%------------- Proposition 2 ----------------------------------------------------------------
%------------- Proposition 2 ----------------------------------------------------------------
%------------- Proposition 2 ----------------------------------------------------------------
{\bf Proposition 4.}  {\it  The PDF and CDF of end-to-end SNR under moderate to strong atmospheric turbulence conditions are derived as}
\begin{align}
	\label{strong_snr_pdf}
	&f_{\Upsilon}(\Upsilon) =   \frac{B_s  B_n}{2\sqrt{\Upsilon_1 \Upsilon}}\sum_{n=1}^N\\
	%
	%-------------
	&\left[
	G^{6,0}_{2,6}\left( B_n' \sqrt{\frac{\Upsilon}{\Upsilon_1}} \bigg|
	\begin{array}{c}
		K,1
		\\
		0,\alpha-1,\beta-1,K-1,\alpha-1,\beta-1 
	\end{array}
	\right)  \right. \nonumber \\
	%--------------
	&-\left.G^{6,0}_{2,6}\left(B_n'' \sqrt{\frac{\Upsilon}{\Upsilon_1}} \bigg|
	\begin{array}{c}
		K,1
		\\
		0,\alpha\!-\!1,\beta\!-\!1,K-1,\alpha-1,\beta-1 
	\end{array}
	\right) \right], \nonumber
\end{align}
{\it and}
\begin{align}
	\label{p2}
	&F_{\Upsilon}(\Upsilon) = B_s  B_n \sqrt{\frac{\Upsilon}{\Upsilon_1}} \sum_{n=1}^N\\
	%
	%-------------
	&\left[
	G^{6,1}_{3,7}\left( B_n' \sqrt{\frac{\Upsilon}{\Upsilon_1}} \bigg|
	\begin{array}{c}
		0,K,1
		\\
		0,\alpha-1,\beta-1,K-1,\alpha-1,\beta-1,-1 
	\end{array}
	\right)  \right. \nonumber \\
	%--------------
	&-\! \left.G^{6,1}_{3,7}\left(\!B_n'' \sqrt{\frac{\Upsilon}{\Upsilon_1}} \bigg|
	\begin{array}{c}
		0,K,1
		\\
		0,\alpha\!-\!1,\beta\!-\!1,K\!-\!1,\alpha\!-\!1,\beta\!-\!1,-\!1 
	\end{array}
	\!\!\right)  \!\right], \nonumber
\end{align}
\begin{IEEEproof}
	Please refer to Appendix \ref{strong_dist}.
\end{IEEEproof}

%
%%%%%%%%%%%%%%%%%%%%%%%%%%%%%%%%%%%%%%%%%%%%%%%%%%%%%%%%%%%%%%%%
%%%%%%%%%%%%%%%%%%%%%%%%%%%%%%%%%%%%%%%%%%%%%%%%%%%%%%%%%%%%%%%% VERSUS W_Z
\begin{figure*}
	\centering
	\subfloat[] {\includegraphics[width=1.7 in]{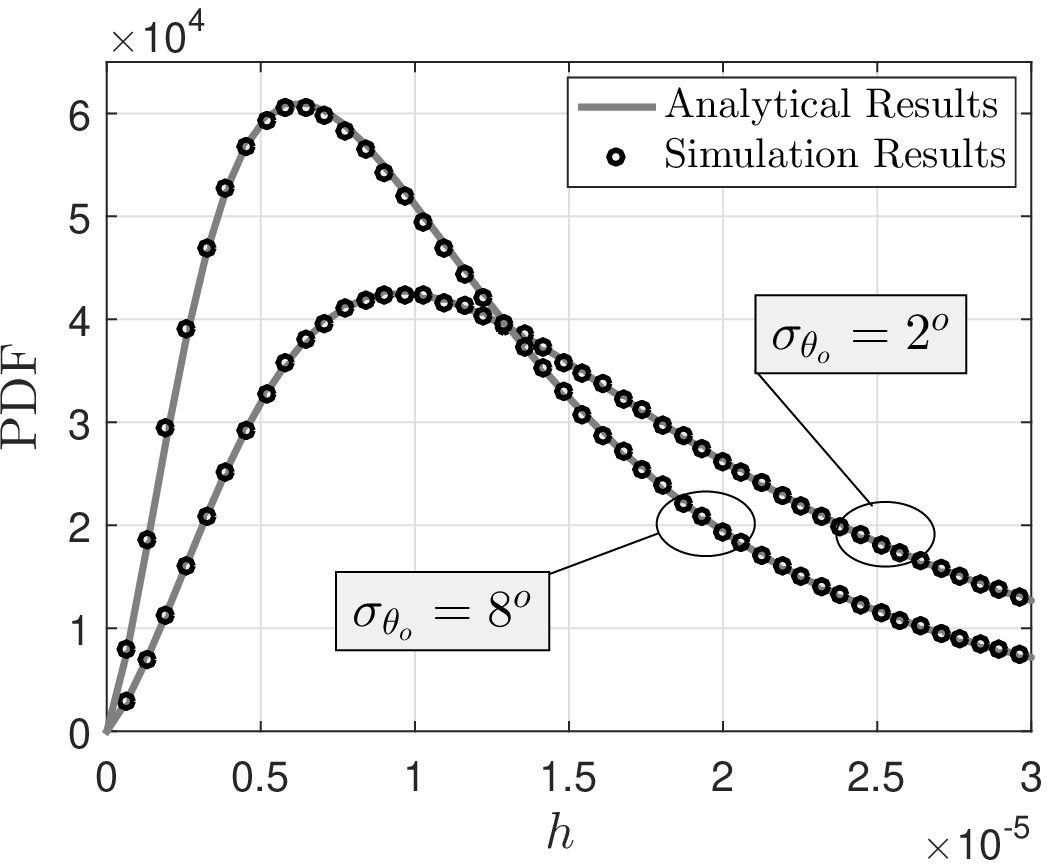}
		\label{cw1}
	}
	\hfill
	\subfloat[] {\includegraphics[width=1.7 in]{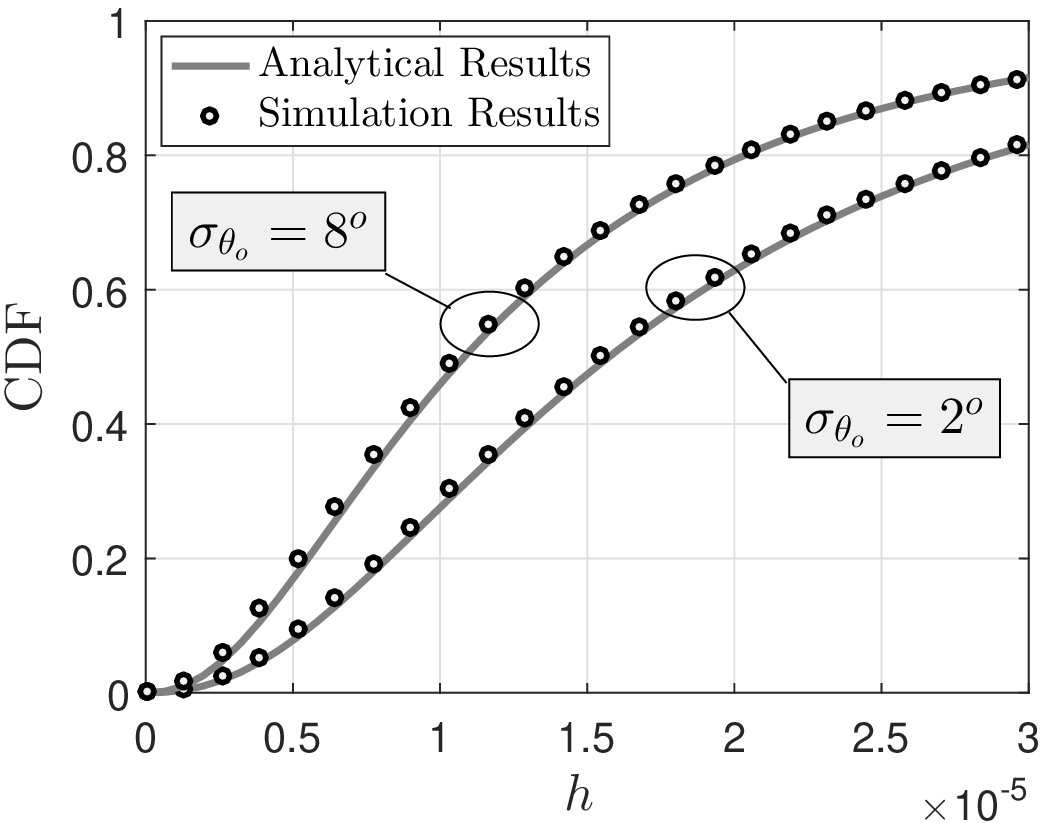}
		\label{cw2}
	}
	\hfill
	\subfloat[] {\includegraphics[width=1.7 in]{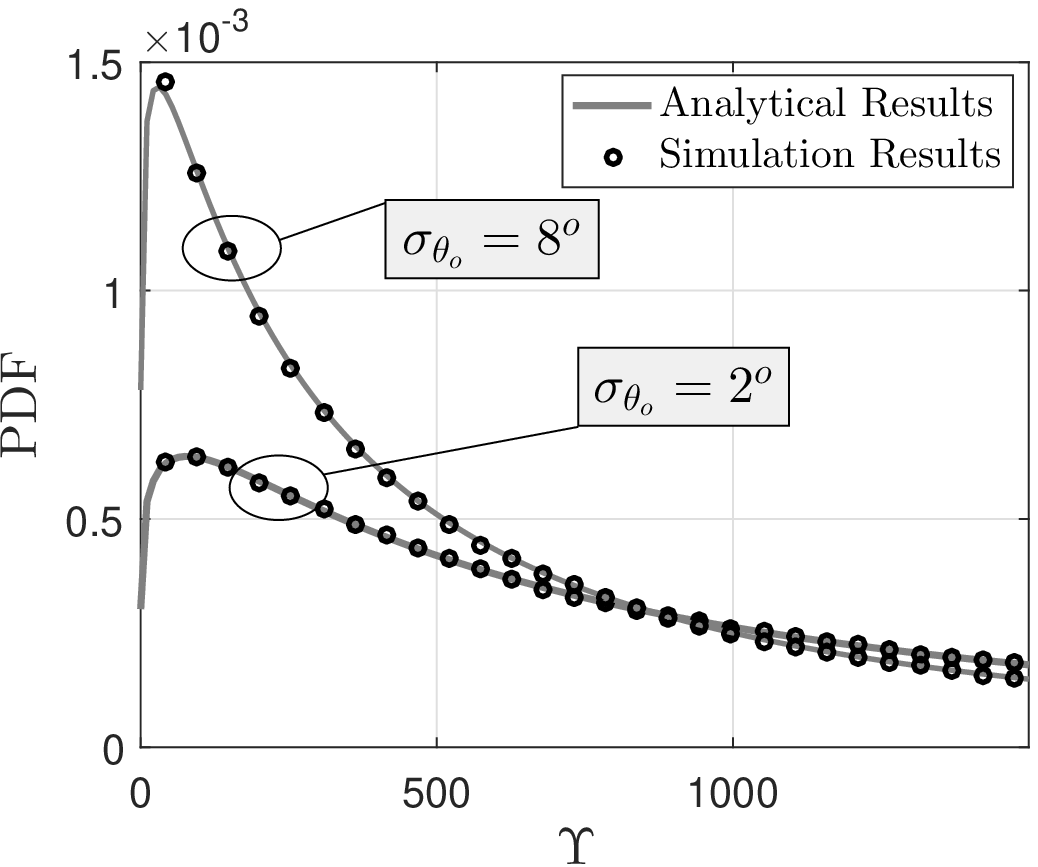}
		\label{cw3}
	}
	\hfill
	\subfloat[] {\includegraphics[width=1.7 in]{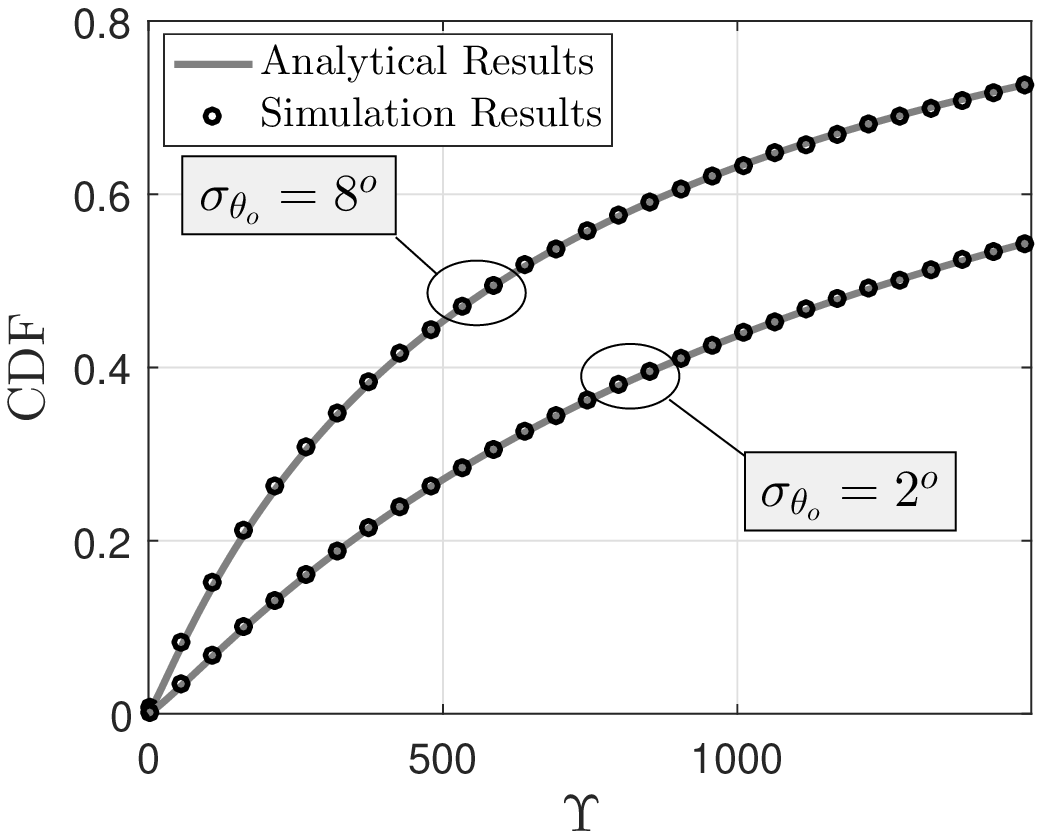}
		\label{cw4}
	}
	\caption{Comparison of the accuracy of the channel models under weak turbulence conditions, when $C_n^2=5\times10^{-15}$, $Z=1000$\,m, and $\sigma_{\theta_e}=100~\micro$rad and for (a) channel PDF given in Theorem 1, (b) CDF of channel coefficients given in Proposition 1, (c) PDF of SNR given in Proposition 2, and (d) CDF of SNR given in Proposition 2.}
	\label{cws}
\end{figure*}
%%%%%%%%%%%%%%%%%%%%%%%%%%%%%%%%%%%%%%%%%%%%%%%%%%%%%%%%%%%%%%%%
%%%%%%%%%%%%%%%%%%%%%%%%%%%%%%%%%%%%%%%%%%%%%%%%%%%%%%%%%%%%%%%%  
%
%
%------------- Proposition 1 ----------------------------------------------------------------
%------------- Proposition 1 ----------------------------------------------------------------
%------------- Proposition 1 ----------------------------------------------------------------
%------------- Proposition 1 ----------------------------------------------------------------
%------------- Proposition 1 ----------------------------------------------------------------
{\bf Proposition 5.}  {\it  For small values of $\sigma_{\theta_o}$, \eqref{st8} and \eqref{cdf_strong} can be simplified respectively as}
\begin{align}
	\label{st9}
	&f_{h}(h) \simeq \frac{  \pi w_z^2 \alpha^2 \beta^2 K  }
	{2A_r (\Gamma(\alpha)\Gamma(\beta))^2 }  \times \\
	%-------------
	&G^{5,0}_{1,5}\left( \frac{ \pi w_z^2\alpha^2 \beta^2 }{2A_r }   h \bigg|
	\begin{array}{c}
		K\\
		\alpha-1, \beta-1, K-1,\alpha-1,\beta-1
	\end{array}
	\right),  \nonumber
\end{align}
and   
\begin{align}
	\label{cdf_aprox_strong}
	&F_{h}(h)  \simeq \frac{  \pi w_z^2 \alpha^2 \beta^2 K   }
	{2A_r (\Gamma(\alpha)\Gamma(\beta))^2 } h  \times \\
	%-------------
	& G^{5,1}_{2,6}\left(\!\! \frac{ \pi w_z^2\alpha^2 \beta^2 }{2A_r } h\bigg|\!\!
	\begin{array}{c}
		0,K\\
		\alpha-1, \beta-1, K-1,\alpha-1,\beta-1,-1
	\end{array}
	\!\!\right) . \nonumber
\end{align}

{\it Also, for small values of $\sigma_{\theta_o}$,  \eqref{strong_snr_pdf} and \eqref{p2} can be simplified respectively as}
\begin{align}
	\label{xp1}
	&f_{\Upsilon}(\Upsilon) \simeq \frac{  \pi w_z^2 \alpha^2 \beta^2 K  }
	{2A_r (\Gamma(\alpha)\Gamma(\beta))^2 } 
	\frac{1}{2\sqrt{\Upsilon_1 \Upsilon}} \times \\
	%-------------
	&G^{5,0}_{1,5}\left( \frac{ \pi w_z^2\alpha^2 \beta^2 }{2A_r }   \sqrt{\frac{\Upsilon}{\Upsilon_1}} \bigg|
	\begin{array}{c}
		K\\
		\alpha\!-\!1, \beta\!-\!1, K\!-\!1,\alpha-1,\beta-1
	\end{array}
	\right),  \nonumber
\end{align}
{\it and}
\begin{align}
	\label{p3}
	&F_{\Upsilon}(\Upsilon)  \simeq \frac{  \pi w_z^2 \alpha^2 \beta^2 K   }
	{2A_r (\Gamma(\alpha)\Gamma(\beta))^2 } \sqrt{\frac{\Upsilon}{\Upsilon_1}}  \times \\
	%-------------
	& G^{5,1}_{2,6}\left(\!\! \frac{ \pi w_z^2\alpha^2 \beta^2  }{2A_r }     \sqrt{\frac{\Upsilon}{\Upsilon_1}}  \bigg|\!\!
	\begin{array}{c}
		0,K\\
		\alpha-1, \beta-1, K-1,\alpha-1,\beta-1,-1
	\end{array}
	\!\!\right) . \nonumber 
\end{align}
%%%%%%%%%%%%%%%%%%%%%%%%%%%%%%%%%%
%%%%%%%%%%%%%%%%%%%%%%%%%%% begin PROOF %%%%%%%%%%%%%%%%%%%%%%%%%%%%%%%%%%%%%%%%%%%%%%%%%%%%%%%%%%%%%%%%%%%%%%%%%%%%%%%%%%%%%%%%%%%%%%%%%%%%%%%%%%%%%%%%%%%%%%%%%%
\begin{IEEEproof}
	Please refer to Appendix \ref{strong_dist}.
\end{IEEEproof}

{In the next section, we show that for UAVs with higher stability, i.e., $\sigma_{\theta_o}<2^o$, the results of Proposition 5 is accurate.}

%
%------------- Proposition 1 ----------------------------------------------------------------
%------------- Proposition 1 ----------------------------------------------------------------
%------------- Proposition 1 ----------------------------------------------------------------
%------------- Proposition 1 ----------------------------------------------------------------
%------------- Proposition 1 ----------------------------------------------------------------
{\bf Proposition 6.}  {\it The closed-form expression for BER of the considered MRR-based FSO communication for OOK modulation over moderate to strong turbulence conditions is derived as}
%
%[12, eq. (06.27.26.0006.01)]
%%%%%%%%%%%%%%%%%%%%%%%%%%%%%%%%%
%%%%%%%%%%%%%%%%%%%%%%%%%%%%%%%%%
\begin{align}
	\label{b5}
	&\mathbb{P}_e = \frac{B_s}{4\sqrt{\pi\Upsilon_1 \Upsilon}}\sum_{n=1}^N   B_n
	\frac{ 2^{3\alpha+2\beta}  }
	{256 \pi^2}  
	\left[
	G^{12,2}_{6,13}\left( \frac{(B_n')^2}{32 \Upsilon_1} \bigg|
	\begin{array}{c}
		<\!a_i\!>
		\\
		<\!b_j\!> 
	\end{array}
	\right)  \right. \nonumber \\
	%--------------
	&-\left.G^{12,2}_{6,13}\left( \frac{(B_n'')^2}{32 \Upsilon_1} \bigg|
	\begin{array}{c}
		<\!a_i\!>
		\\
		<\!b_j\!>
	\end{array}
	\right) \right]
\end{align}
where
\begin{align}
	%\label{sr9}
	\left\{
	\begin{array}{ll}
		&\!\!\!\!\!\!\!<\!a_i\!> = 0,\frac{1}{2}, \frac{K}{2}, \frac{K+1}{2}, \frac{1}{2}, 1 \nonumber \\
		%---------
		&\!\!\!\!\!\!\!<\!b_j\!> = 0,\frac{1}{2},\frac{\alpha-1}{2}, \frac{\alpha}{2},      
		\frac{\beta-1}{2}, \frac{\beta}{2}, \frac{K-1}{2}, \frac{K}{2},
		\frac{\alpha-1}{2}, \frac{\alpha}{2}, \frac{\beta-1}{2}, \frac{\beta}{2},    -1 .
	\end{array} 
	\right.  \nonumber 
\end{align}

%%%%%%%%%%%%%%%%%%%%%%%%%%%%%%%%%%
%%%%%%%%%%%%%%%%%%%%%%%%%%% begin PROOF %%%%%%%%%%%%%%%%%%%%%%%%%%%%%%%%%%%%%%%%%%%%%%%%%%%%%%%%%%%%%%%%%%%%%%%%%%%%%%%%%%%%%%%%%%%%%%%%%%%%%%%%%%%%%%%%%%%%%%%%%%
\begin{IEEEproof}
	Please refer to Appendix \ref{strong_BER}.
\end{IEEEproof}

%%%%%%%%%%%%%%%%%%%%%%%%%%%%%%%%%%%%%%%%%%%%%%%%%%%%%%%%%%%%%%%%
\begin{table}
	\caption{System Parameters for Simulations.} % title of Table
	\centering % used for centering table
	\begin{tabular}{l c c} % centered columns (3 columns)
		\hline\hline \\[-1.2ex]%inserts double horizontal lines\\
		{\bf Description} & {\bf Parameter} & {\bf Setting} \\ [.5ex] % inserts table
		%heading
		\hline\hline \\[-1.2ex]% inserts single horizontal line
		PD responsivity   &$R$ & 0.8 A/W \\
		SNR threshold   &$\Upsilon_\textrm{th}$ &  5 dB \\
		Radius of GS aperture &$r_g$ & 8 cm \\
		MRR effective area                        & $ A_r $           &0.5-4 $\textrm{cm}^2$  \\[1ex] 
		Link length             & $ Z $         & 500-1500 m \\[1ex]
	    SD of UAV orientation fluctuations & $\sigma_{\theta_o}$ & 1-8 in degree \\[1ex]
	    SD of tracking angle errors & $\sigma_{\theta_e}$ & 50-400 $\micro$rad \\[1ex]
		Wavelength                      &$ \lambda $      &$ 1550$ nm   \\
		Divergence angle   & $\theta_\textrm{div}$ &         0.1-2 mrad \\
		Beamwidth at the Rx       &$w_z$ & 0.1-2 m \\
		Transmit power                       &$P_t$        & 0-30  dBm\\ 
		Noise variance       &$\sigma_n^2$           & -11 dBm \\
		Number of sectors &$N$   & 8  \\
		Channel Loss                 & $ h_{l_{gu}}~\&~ h_{l_{ug}}$           & 0.7 \\
		Refractive-index structure  & $C_n^2$ &  $10^{-14}-10^{-13}$ \\
		for moderate-to-strong turbulence               &  &  \\
		Refractive-index structure  & $C_n^2$ &  $10^{-15}-10^{-14}$ \\
		for weak-to-moderate turbulence               &  &  \\
		\hline \hline              
	\end{tabular}
	\label{I} % is used to refer this table in the text
\end{table}
%%%%%%%%%%%%%%%%%%%%%%%%%%%%%%%%%%%%%%%%%%%%%%%%%%
%%%%%%%%%%%%%%%%%%%%%%%%%%%%%%%%%%%%%%%%%%%%%%%%%%

%
%
%%%%%%%%%%%%%%%%%%%%%%%%%%%%%%%%%%%%%%%%%%%%%%%%%%%%%%%%%%%%%%%%
%%%%%%%%%%%%%%%%%%%%%%%%%%%%%%%%%%%%%%%%%%%%%%%%%%%%%%%%%%%%%%%%
\begin{figure}
	\begin{center}
		\includegraphics[width=3.2 in]{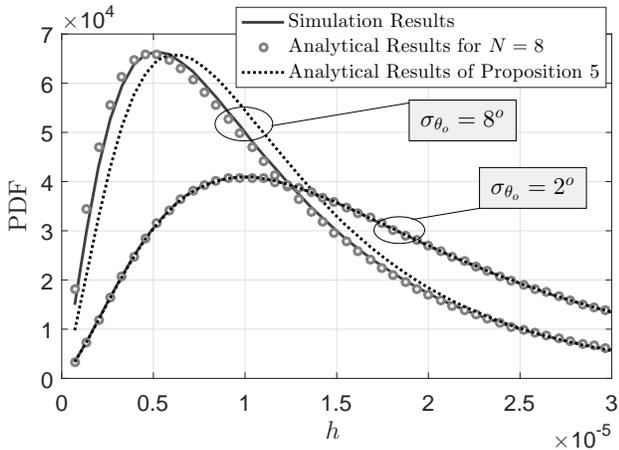}
		\caption{ Comparison of the accuracy of the channel models provided in Theorem 2 and Proposition 5 under moderate to strong turbulence conditions, when $Z=1000$\,m, and $\sigma_{\theta_e}=90~\micro$rad.}
		\label{gg_fig}
	\end{center}
\end{figure}
%%%%%%%%%%%%%%%%%%%%%%%%%%%%%%%%%%%%%%%%%%%%%%%%%%%%%%%%%%%%%%%%
%%%%%%%%%%%%%%%%%%%%%%%%%%%%%%%%%%%%%%%%%%%%%%%%%%%%%%%%%%%%%%%%
%
%

%
%
%%%%%%%%%%%%%%%%%%%%%%%%%%%%%%%%%%%%%%%%%%%%%%%%%%%%%%%%%%%%%%%%
%%%%%%%%%%%%%%%%%%%%%%%%%%%%%%%%%%%%%%%%%%%%%%%%%%%%%%%%%%%%%%%%
\begin{figure}
	\begin{center}
		\includegraphics[width=3.2 in]{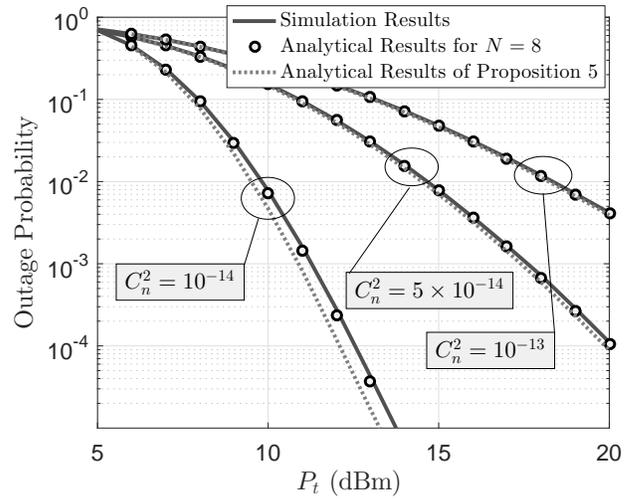}
		\caption{ Outage probability versus $P_t$ for $Z=1000$, $\sigma_{\theta_o}=6^o$, $\sigma_{\theta_e}=100~\micro$rad, and three different values of $C_n^2=10^{-14}$, $5\times10^{-14}$, and $10^{-13}$.  }
		\label{gg_out}
	\end{center}
\end{figure}
%%%%%%%%%%%%%%%%%%%%%%%%%%%%%%%%%%%%%%%%%%%%%%%%%%%%%%%%%%%%%%%%
%%%%%%%%%%%%%%%%%%%%%%%%%%%%%%%%%%%%%%%%%%%%%%%%%%%%%%%%%%%%%%%%
%
%

%
%
%%%%%%%%%%%%%%%%%%%%%%%%%%%%%%%%%%%%%%%%%%%%%%%%%%%%%%%%%%%%%%%%
%%%%%%%%%%%%%%%%%%%%%%%%%%%%%%%%%%%%%%%%%%%%%%%%%%%%%%%%%%%%%%%%
\begin{figure}[!]
	\begin{center}
		\includegraphics[width=3.2 in]{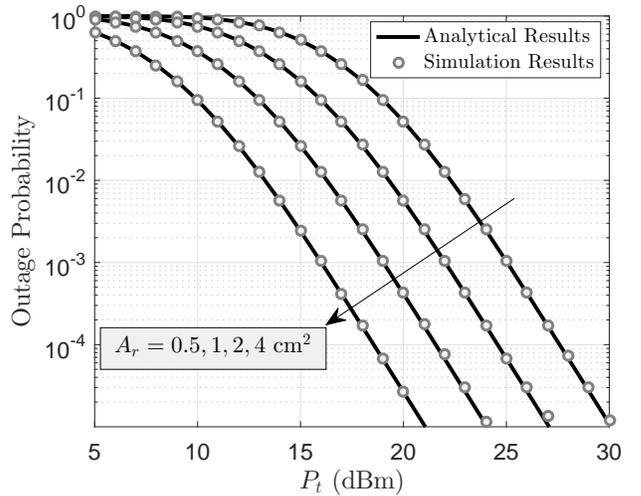}
		\caption{ Outage probability versus $P_t$ for $Z=1000$, $\sigma_{\theta_o}=6^o$, $\sigma_{\theta_e}=100~\micro$rad, $w_z=40$ cm, and different values of effective aperture area $A_r$.  }
		\label{differ_Ar}
	\end{center}
\end{figure}
%%%%%%%%%%%%%%%%%%%%%%%%%%%%%%%%%%%%%%%%%%%%%%%%%%%%%%%%%%%%%%%%
%%%%%%%%%%%%%%%%%%%%%%%%%%%%%%%%%%%%%%%%%%%%%%%%%%%%%%%%%%%%%%%%
%
%
%BER_log_div_04_Ar1_differ_Z

%-----------------------------------
%-----------------------------------
\section{Numerical Results}
%-----------------------------------
%-----------------------------------
In this section, first, we utilize computer simulations to verify the accuracy of our proposed analytical channel models for MRR-based FSO system. 
Second, the performance of the considered system is studied in terms of outage probability and BER.
%.
The main considered parameters for the simulation results are summarized in Table \ref{I}, mostly adopted from the standard values for system parameters in \cite{ghassemlooy2019optical}.

For evaluation of analytical channel models provided in Section III, we perform Monte-Carlo simulations. The details of the simulation process are described as follows. 
For a given $\sigma_{\theta_o}$, we generate $5\times10^7$ independent RVs $\theta_{xz}$, $\theta_{xy}$ and $\theta_{yz}$. 
Then, based on \eqref{g2}, we generate $5\times10^7$ independent coefficients of $h_{MRR_n}$ for $n\in\{xy,xz,yz\}$. Now, using \eqref{g1}, $5\times10^7$ independent coefficients of $h_{MRR}$ are generated.
Moreover, for a given $\sigma_{\theta_e}$, we generate $5\times10^7$ independent RVs $\theta_{ex}$ and $\theta_{ey}$. Then, using the generated RVs $\theta_{ex}$ and $\theta_{ey}$, we generate $5\times10^7$ independent coefficients of $h_{pu}$ from \eqref{po1} and \eqref{ss2}. 
For a given $C_n^2<10^{-14}$, we also generate $5\times10^7$ independent coefficients of $h_{a_{gu}}$ and $h_{a_{ug}}$ which have log-normal distribution as given in \eqref{ss9}. For a given $C_n^2>10^{-14}$, we generate $5\times10^7$ independent coefficients of $h_a$ which have GG distribution as given in \eqref{ss8}. We then obtain $5\times10^7$ independent values of UAV-based optical channel coefficients based on \eqref{s}. Finally, we find the channel distribution diagrams.
%
%It is worth mentioning that, for each state of simulation, we perform independent runs in MATLAB which takes about 20 minutes of processing time (Intel Core i7 Processors, 8 GB RAM). On the other hand, by using our proposed analytical-based methods proposed in Section III, the channel can be easily modeled in less than a second which is extremely faster than employing simulation-based methods.

The accuracy of the proposed channel models under weak turbulence conditions is evaluated in Fig. \ref{cws} for two different values of the UAV's instability parameters $\sigma_{\theta_o}=2^o$ and $8^o$. In particular, we corroborate the accuracy of PDF of instantaneous channel coefficients and end-to-end SNR (respectively provided in Theorem 1 and Proposition 2) in Figs. \ref{cw1} and \ref{cw3}, respectively, and we corroborate the accuracy of CDF of instantaneous channel coefficients and end-to-end SNR (respectively provided in Propositions 1 and 2) in Figs. \ref{cw2} and \ref{cw4}, respectively. The simulation results clearly confirm the accuracy of analytical models under weak turbulence conditions. 
In Fig. \ref{gg_fig}, we also investigate the accuracy of PDF of instantaneous channel coefficients under strong turbulence conditions for two different values of the UAV's instability parameters $\sigma_{\theta_o}=2^o$ and $8^o$.
From Theorem 2, the parameter $N$ impacts on the validity of channel PDF.
The variable $N$ is used for approximating the distribution of RV $h_{MRR}$.
The optimal value for $N$ is its minimum value that satisfies a
predefined accuracy.
The results of Fig. \ref{gg_fig} clearly show that the analytical channel model derived in Theorem 2 with $N=8$ is valid for all conditions. 
In Proposition 5, we also propose a more tractable closed-form channel model. As we observe, the analytical channel model
derived in proposition 5 is accurate for more stable UAVs with $\sigma_{\theta_o}<2^o$.

The effect of atmospheric turbulence conditions on optical link performance can be characterized by the index of refraction structure parameter $C_n^2$. 
In Fig. \ref{gg_out}, outage probability is plotted for $Z=1000$, $\sigma_{\theta_o}=6^o$, $\sigma_{\theta_e}=100~\micro$rad, and a wide range of $C_n^2$ i.e., $C_n^2=10^{-14}$, $5\times10^{-14}$, and $10^{-13}$. The results of Fig. \ref{gg_out} clearly show the impairments caused by atmospheric turbulence on the performance of MRR-based FSO system. The effect of atmospheric turbulence on the considered MRR-based FSO system is more severe than the conventional FSO system because in MRR FSO system, the atmospheric turbulence affects both GS-to-UAV and UAV-to-GS links characterized by $h_{a_{gu}}$ and $h_{a_{ug}}$, respectively.

The effective area of MRR denoted by $A_r$ is another important parameter that plays a key role in the link budget. In Fig. \ref{differ_Ar}, we investigate the effect of $A_r$ on the performance of MRR FSO system. A larger value of $A_r$ lets the MRR to collect more optical power and thus, improves the link budget which leads to a lower outage probability. 
However, in a practical implementation, we are not allowed to use large values for $A_r$ because the switching rate of MRR modulator is inversely proportional to $A_r$. Accordingly, in practice, the active area of MRR is usually selected by a trade-off between link budget and desired data rate.

%
%
%%%%%%%%%%%%%%%%%%%%%%%%%%%%%%%%%%%%%%%%%%%%%%%%%%%%%%%%%%%%%%%%
%%%%%%%%%%%%%%%%%%%%%%%%%%%%%%%%%%%%%%%%%%%%%%%%%%%%%%%%%%%%%%%%
\begin{figure}
	\begin{center}
		\includegraphics[width=3.2 in]{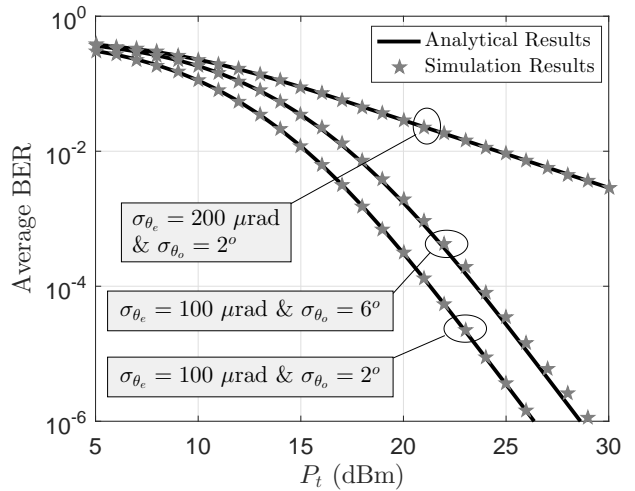}
		\caption{ Average BER of MRR-based FSO system for $w_z=30$\,cm, $Z=1000$ m,
			and different values of $\sigma_{\theta_e}$ and $\sigma_{\theta_o}$.   }
		\label{df}
	\end{center}
\end{figure}
%%%%%%%%%%%%%%%%%%%%%%%%%%%%%%%%%%%%%%%%%%%%%%%%%%%%%%%%%%%%%%%%
%%%%%%%%%%%%%%%%%%%%%%%%%%%%%%%%%%%%%%%%%%%%%%%%%%%%%%%%%%%%%%%%
%
%Outage_vesus_wz_differ_Z_PtdBm20

Other main parameters for a UAV-based MRR system are the SD of tracking system errors and the SD of UAV orientation fluctuations which are denoted by $\sigma_{\theta_e}$ and $\sigma_{\theta_o}$, respectively. These parameters have a significant impact on system performance and related link budget. To get a better insight, in Fig. \ref{df}, the BER of the considered system is plotted for different values of $\sigma_{\theta_e}$ and $\sigma_{\theta_o}$.  
From the results of Fig. \ref{df}, at the target BER $\mathbb{P}_e=10^{-6}$ and $\sigma_{\theta_e}=100$\,$\micro$rad, the considered system requires approximately 3 dB more transmit power to compensate the degrading effect of increasing UAV's instability from $\sigma_{\theta_o}=2^o$ to $6^o$. However, for $\sigma_{\theta_o}=2^o$, by increasing tracking system error from $\sigma_{\theta_e}=100$ to 200\,$\micro$rad, the performance significantly degrades.
As a result, for an MRR based FSO system, the parameter $\sigma_{\theta_e}$ has a greater impact on performance of MRR-based FSO system with respect to $\sigma_{\theta_o}$.
From the results of \cite{dabiri2019optimal,khankalantary2020ber}, for a short link UAV-based FSO system, the tolerable UAVs orientation fluctuations is in the order of several mrads and for a long link, is in the order of $\micro$rad. 
However, for the small UAVs with low power and payload limitations (that it is not possible to employ stabilizers) it is not possible to achieve such a UAV's angular stability.
One of the main advantages of employing MRR is to compensate of UAV's orientation fluctuations, especially, for the small UAVs. From the results of Fig. \ref{df}, the considered MRR-based system has an acceptable performance even for high UAV's angular fluctuations equal to $\sigma_{\theta_o}=6^o=104.7$\,mrad.
However, due to the limited data rate, we are forced to use a smaller $A_r$ that makes this MRR-based system more sensitive to tracking system errors with respect to the conventional FSO systems.
The results of Fig. \ref{df} clearly confirm the aforementioned points.

%
%
%%%%%%%%%%%%%%%%%%%%%%%%%%%%%%%%%%%%%%%%%%%%%%%%%%%%%%%%%%%%%%%%
%%%%%%%%%%%%%%%%%%%%%%%%%%%%%%%%%%%%%%%%%%%%%%%%%%%%%%%%%%%%%%%%
\begin{figure}
	\begin{center}
		\includegraphics[width=3.2 in]{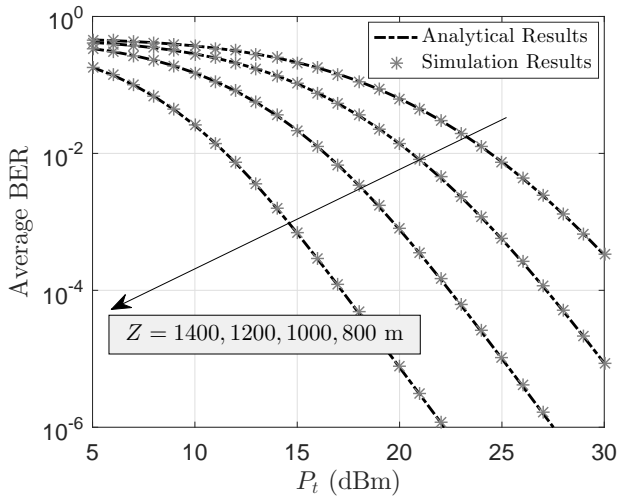}
		\caption{ Average BER versus $P_t$ for $\theta_{div}=0.4$ mrad, $A_r=1\,\textrm{cm}^2$, and different link lengths $Z=800$, $1000$, $1200$, and $1400$.  }
		\label{differ_z}
	\end{center}
\end{figure}
%%%%%%%%%%%%%%%%%%%%%%%%%%%%%%%%%%%%%%%%%%%%%%%%%%%%%%%%%%%%%%%%
%%%%%%%%%%%%%%%%%%%%%%%%%%%%%%%%%%%%%%%%%%%%%%%%%%%%%%%%%%%%%%%%
%
%BER_compare_tet_UAV_and_Ground

\begin{figure}
	\begin{center}
		\includegraphics[width=3.2 in]{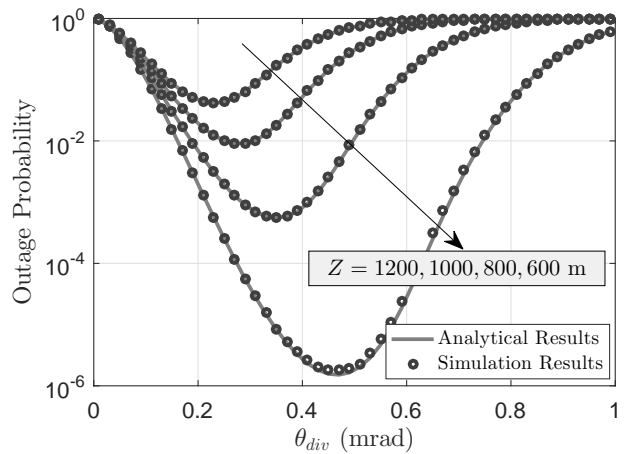}
		\caption{ Outage probability versus $\theta_{div}$ for $P_t=20$\,dBm, $\sigma_{\theta_o}=5^o$, $\sigma_{\theta_e}=100$\,$\micro$rad, and different values of link length.  }
		\label{df2}
	\end{center}
\end{figure}

Now, the impact of link length is evaluated on the performance of the considered MRR-based FSO system in Figs. \ref{differ_z} and \ref{df2}.
In Fig. \ref{differ_z}, the BER performance is evaluated for different values of link lengths $Z=800$, $1000$, $1200$, and $1400$.
As expected, by increasing link length, the performance degrades, significantly. 
However, it should be noted that the beamwidth at the MRR changes by varying link length which changes the distribution of RV $h_{pu}$ and end-to-end SNR. 
Accordingly, for any given link length, we must find an optimal beamwidth to achieve minimum outage probability and/or BER. 
The beamwidth is tuned by divergence angle at the GS node. Thus, for any given link length, in the considered MRR-based FSO system, finding and tuning an optimal value for divergence angle is very important. 
To get a better insight, in Fig. \ref{df2}, the outage probability is plotted versus $\theta_{div}$ for different values of link length. The results of Fig. \ref{df2} clearly shows by varying link length, the optimal value for $\theta_{div}$ changes and confirm the importance of finding optimal value for $\theta_{div}$ when the link length is varied.

\begin{figure}
	\begin{center}
		\includegraphics[width=3.2 in]{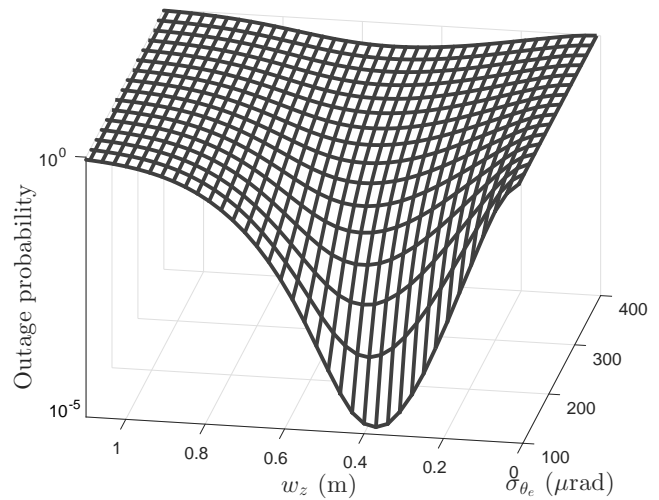}
		\caption{ Outage probability versus $\sigma_{\theta_e}$ and $w_z$ for $P_t=25$\,dBm. }
		\label{df3}
	\end{center}
\end{figure}

In addition to the link length, any changing in other parameters such as $\sigma_{\theta_e}$ can change the optimal value for beamwidth.
In Fig. \ref{df3}, outage probability is depicted versus $\sigma_{\theta_e}$ and $w_z$. The results of Fig. \ref{df3} shows that any 
increase in the SD of tracking system errors causes an increase in the optimal value for beamwidth. 
This can be justified since by increasing $\sigma_{\theta_e}$, the beamwidth must be increased to compensate the fluctuations of Gaussian beam footprint at the MRR. However, any increase in beamwidth increases the geometrical loss. Therefore, for any given $\sigma_{\theta_e}$, the optimal value for $w_z$ can be obtained by a trade-off between the strength of Gaussian beam footprint fluctuations and geometrical loss.

%
%
%%%%%%%%%%%%%%%%%%%%%%%%%%%%%%%%%%%%%%%%%%%%%%%%%%%%%%%%%%%%%%%%
%%%%%%%%%%%%%%%%%%%%%%%%%%%%%%%%%%%%%%%%%%%%%%%%%%%%%%%%%%%%%%%%
%%%%%%%%%%%%%%%%%%%%%%%%%%%%%%%%%%%%%%%%%%%%%%%%%%%%%%%%%%%%%%%%
%%%%%%%%%%%%%%%%%%%%%%%%%%%%%%%%%%%%%%%%%%%%%%%%%%%%%%%%%%%%%%%%
%
%threeD_Pt25dBm

%
%
%%%%%%%%%%%%%%%%%%%%%%%%%%%%%%%%%%%%%%%%%%%%%%%%%%%%%%%%%%%%%%%%
%%%%%%%%%%%%%%%%%%%%%%%%%%%%%%%%%%%%%%%%%%%%%%%%%%%%%%%%%%%%%%%%
%%%%%%%%%%%%%%%%%%%%%%%%%%%%%%%%%%%%%%%%%%%%%%%%%%%%%%%%%%%%%%%%
%%%%%%%%%%%%%%%%%%%%%%%%%%%%%%%%%%%%%%%%%%%%%%%%%%%%%%%%%%%%%%%%
%
%

%-----------------------------------
%-----------------------------------
\section{Conclusion and Future Road Map}
%-----------------------------------
%-----------------------------------
In this paper, we have studied the performance of UAV-based FSO link when UAV is equipped with MRR.
Accordingly, we have characterized the MRR-based UAV FSO channel by taking into account tracking system errors along with UAV's orientation fluctuations, link length, UAV's height, optical beam divergence angle, effective area of MRR, atmospheric turbulence and optical channel loss in the double-pass channels.
To enable effective performance analysis, we have derived the tractable and closed-form expressions for PDF of end-to-end SNR, outage probability and BER of the considered system under both weak-to-moderate and moderate-to-strong  atmospheric turbulence conditions. 
We have then verified the accuracy of analytical models by employing Monte Carlo simulations. 
%
%As shown, unlike  the conventional FSO systems, optimal design of a UAV-based MRR FSO system is very important and any change in the parameters (such as link length, SD of tracking system errors, target BER, desired data rate, etc.) affect the optimal values of other parameters. 
Our results reveal that any change in the parameters (such as link length, SD of tracking system errors, target BER, desired data rate, etc.) affect the optimal values of other parameters. 
For MRR-based FSO deployments, the proposed analytical methods will assist researchers to easily analyze and design of such systems without performing any time-consuming simulations.

Thanks to their high internal gain, avalanche photo-detector (APD) can improve SNR capability, as compared with PIN-based receivers. However, in such APD-based receivers, shot noise is mostly dominant. The variance of shot noise depends on the received optical signal intensity and thus, system analysis becomes more complex and can be considered as a future work.

%Our analytical results have made it possible to find the optimal antenna directivity gain for designing a reliable UAV-based mmW communications  under different levels of stability of UAVs without resorting to time-consuming simulations.

%\newpage
%
%
%%---- APPENDIX ---------------------------------------------------------------------
%%---- APPENDIX ---------------------------------------------------------------------
%%---- APPENDIX ---------------------------------------------------------------------
%
%
%%---- APPENDIX ---------------------------------------------------------------------
%%---- APPENDIX ---------------------------------------------------------------------
%%---- APPENDIX ---------------------------------------------------------------------
%
%
%%---- APPENDIX ---------------------------------------------------------------------
%%---- APPENDIX ---------------------------------------------------------------------
%%---- APPENDIX ---------------------------------------------------------------------

\appendices

%
%%%%%%%%%%%%%%%%%%%%%%%%%%%%%%%%%%%%%%%%%%%%%%%%%%%%%%%%%%%%%%%%
%%%%%%%%%%%%%%%%%%%%%%%%%%%%%%%%%%%%%%%%%%%%%%%%%%%%%%%%%%%%%%%%
\begin{figure}
	\begin{center}
		\includegraphics[width=3.3 in]{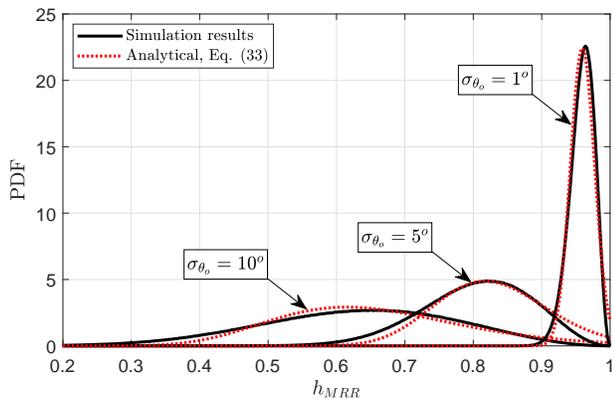}
		\caption{Comparison of the accuracy of the provided PDF for the RV $h_{MRR}$ in \eqref{sr2} with actual distribution under under different SD of UAV orientations.}
		\label{n4}
	\end{center}
\end{figure}
%%%%%%%%%%%%%%%%%%%%%%%%%%%%%%%%%%%%%%%%%%%%%%%%%%%%%%%%%%%%%%%%
%%%%%%%%%%%%%%%%%%%%%%%%%%%%%%%%%%%%%%%%%%%%%%%%%%%%%%%%%%%%%%%%
%channel_coefficients

%%-------------------------------------------------
%%-------------------------------------------------
\section{The Proof of Theorem 1}
\label{weak_dist}
We reformulate \eqref{s} as
\begin{align}
\label{sr}
h = h_{c}  h_L h_{p_u},
\end{align}
where $h_{c}=h_{l_{ug}} h_{l_{gu}} h_{p_g}$ is a constant coefficient and $h_{L}=h_{a_{ug}} h_{a_{gu}} h_{MRR}$.
After an exhaustive search over the distribution of $h_{MRR}$ denoted by $f_{h_{MRR}}(h_{MRR})$, we obtain an approximate mathematical function for $f_{h_{MRR}}(h_{MRR})$ as
\begin{align} % \mu_{MRR}  \sigma_{MRR}^2 
\label{sr2}
&f_{h_{MRR}}(h_{MRR}) \simeq \frac{1}{h_{MRR}\sqrt{2\pi \ln\left(1+\frac{\sigma_{MRR}^2}{\mu_{MRR}^2}\right)}} \\
%-----------
&~~~\times\exp\left(-\frac{\left[\ln\left(\frac{\sqrt{\mu_{MRR}^2+\sigma_{MRR}^2}}{\mu_{MRR}^2} h_{MRR}\right)\right]^2}
{2\ln\left(1+\frac{\sigma_{MRR}^2}{\mu_{MRR}^2}\right)}\right).	\nonumber
\end{align}
To find a better view about the accuracy of the approximated PDF in 32, the accuracy of the approximated PDF for $h_{MRR}$ is compared with actual distribution in Fig. \ref{n4} under different SD of UAV's orientations.
As shown in Fig. \ref{n4}, the results obtained from \eqref{sr2} is very close to the actual values of the PDF of $h_{MRR}$ for $\sigma_{\theta_o}=1^o$. As $\sigma_{\theta_o}$ increases, it is observed that the results of \eqref{sr2} deviate slightly from the actual values, and for $\sigma_{\theta_o}=10^o$ this deviation increases. As will be shown in the simulation section, for the interval $\sigma_{\theta_o}<8^o$, the end-to-end channel distribution function is well close to the value obtained from the Monte-Carlo simulations, and for interval $\sigma_{\theta_o}>8^o$, we will see an offset between the analytical and the Monte-Carlo results. This offset is mainly caused by \eqref{sr2}.
From \eqref{ss9} and \eqref{sr2}, after some derivations, we obtain
\begin{align}
\label{sr3}
&f_{h_L}(h_L) = \frac{1}{h_L\sqrt{2\pi \left[\ln\left(1+\frac{\sigma_{MRR}^2}{\mu_{MRR}^2}\right) + 8\sigma_L^2 \right]}} \\
%-----------
&~~~\times\exp\left(-\frac{\left[\ln\left(\frac{\sqrt{\mu_{MRR}^2+\sigma_{MRR}^2}}{\mu_{MRR}^2} h_L\right)+4\sigma_L^2\right]^2}
{2\left[\ln\left(1+\frac{\sigma_{MRR}^2}{\mu_{MRR}^2}\right) + 8\sigma_L^2 \right]}\right).	\nonumber
\end{align}
%----------
%
% In practice, $\sigma_{\theta_e}$ is in the order of a few hundred $\micro$rad to a few mrad.
As shown in simulation results, for a reliable communication, the SD of tracking system errors characterized by $\theta_{ex}$ small (less than a few mrad). Under such tracking errors, \eqref{po1} can be simplified as
\begin{align}
	\label{sr4}
	d_{px} \simeq Z\theta_{ex},	~~~~ \&~~~~ d_{py} \simeq Z\theta_{ey}.
\end{align}
From \eqref{sr4}, the distribution of $d_p=\sqrt{d_{px}^2+d_{py}^2}$ well approximated as
\begin{align}
	\label{sr5}
f_{d_{p}}(d_{p}) = \frac{d_p}{Z^2 \sigma_{\theta_e}^2} \exp\left(-\frac{d_p^2}{2Z^2 \sigma_{\theta_e}^2} \right),~~~~~~d_p\geq0.	
\end{align}
From \eqref{ss3} and \eqref{sr5}, we obtain 
\begin{align}
\label{sr6}
f_{h_{p_u}}(h_{p_u}) = K \left( \frac{\pi w_z^2}{2 A_r}\right)^K h_{p_u}^{K-1},~~~ 0< h_{p_u} \leq \frac{2 A_r}{\pi w_z^2},	
\end{align}
where $K=\frac{w_z^2}{Z^2 \sigma_{\theta_e}^2}$. %, and $A_0 = \frac{2 A_r}{\pi w_z^2}$.
From \eqref{sr}, \eqref{sr3}, and \eqref{sr6}, we have
\begin{align}
\label{sr7}
&f_h(h) = \frac{K(2 A_r h_c)^{-K} (\pi w_z^2)^K  h^{K-1}  }
{\sqrt{ 2\pi \left[\ln\left(1+\frac{\sigma_{MRR}^2}{\mu_{MRR}^2}\right) + 8\sigma_L^2 \right]}}          
\int_{\frac{\pi w_z^2 h}{2 A_r h_c}}^\infty   
h_L^{-K-1}  \nonumber \\
&\times\exp\left(-\frac{\left[\ln\left(\frac{\sqrt{\mu_{MRR}^2+\sigma_{MRR}^2}}{\mu_{MRR}^2} h_L\right)+4\sigma_L^2\right]^2}
{2\left[\ln\left(1+\frac{\sigma_{MRR}^2}{\mu_{MRR}^2}\right) + 8\sigma_L^2 \right]}\right)        dh_L.   
\end{align}
Finally, applying a change of variable $x = \ln\left(\frac{\sqrt{\mu_{MRR}^2+\sigma_{MRR}^2}}{\mu_{MRR}^2} h_L\right)$ and using \cite[eq. (2.33)]{izrail1996table}, the closed-form channel distribution is derived in \eqref{sr8}.

%----------- end --------------------------------------------
%----------- end --------------------------------------------

%%-------------------------------------------------
%%-------------------------------------------------
\section{The Proof of Proposition {3}}
\label{weak_BER}
Let $p_0$ and $p_1$ denote the a priori probability of transmission bits ``0'' and ``1'', respectively.
The BER of intensity modulated direct detection with on-off keying (OOK) signaling is given by $\mathbb{P}_e =p_0 \mathbb{P}_{e|0} +  p_1 \mathbb{P}_{e|1}$ where $\mathbb{P}_{e|0}$ and $\mathbb{P}_{e|1}$ denote the conditional bit error probabilities when the transmitted bit is ``0'' and ``1'', respectively. Considering also that $p_0=p_1$  and $\mathbb{P}_{e|0}=\mathbb{P}_{e|1}$, the BER is derived as \cite{simon2005digital}
\begin{align}
\label{a1}
\mathbb{P}_e = \int_0^\infty  Q\left( \sqrt{\Upsilon}\right) f_\Upsilon(\Upsilon) ~\textrm{d} \Upsilon.	
\end{align}
Substituting \eqref{sr8} in \eqref{a1} and using a series expansion \cite[eq. (06.25.06.0002.01)]{wolfram}, $\mathbb{P}_e$ is well approximated as
\begin{align}
	\label{a2}
	\mathbb{P}_e &\simeq \frac{C_4}{4  \Upsilon_1^{K/2}}    \left(  
	\int_0^{\Upsilon_\textrm{max}}       
	\Upsilon^{\frac{K-2}{2}}    Q\left( \frac{\ln(\Upsilon) - \ln(\Upsilon_1) + 2C_5}{2\sqrt{C_1}}   \right)  \textrm{d}\Upsilon \right. \nonumber\\
	%--------------------           
	&~~~-\frac{2}{\sqrt{\pi}}\sum_{m=0}^M  \frac{(-1)^m }   {m!(2m+1)2^{m+\frac12}  }  \int_0^{\Upsilon_\textrm{max}}   
	\Upsilon^{\frac{2m+K-1}{2}}   \nonumber \\
	%-------------------- 
	&\left.\textcolor{white}{\int_0^{\Upsilon_\textrm{max}}}   \!\!\!\!\!\!  \!\!\!\!\! 
	\times  Q\left( \frac{\ln(\Upsilon) - \ln(\Upsilon_1) + 2C_5}{2\sqrt{C_1}}   \right)  \textrm{d}\Upsilon \right).	
\end{align}
where $M=20$ and $\Upsilon_\textrm{max}=4$.
%The accuracy of approximated BER provided in \eqref{a2} increases by increasing $M$ and $\Upsilon_\textrm{max}$. Notice that the term of $\Upsilon^{\frac{2m+K-1}{2}}$ in \eqref{a2} limits to infinity for large values of $M$ and $\Upsilon_\textrm{max}$ and due to the computational limitations of many existing software, we are not allowed to use any large values for terms $M$ and $\Upsilon_\textrm{max}$. After an exhaustive search, we found that the BER is obtained with good accuracy for $M=20$ and $\Upsilon_\textrm{max}=4$. 
%
%Using the identity \cite[eq. (06.27.21.0011.01)]{wolfram}
In the following derivation, we use an integral identity \cite[eq. (06.27.21.0011.01)]{wolfram}
\begin{align}
\label{a3}
\int e^{b x} \textrm{erfc}(ax) = \frac{1}{b}
\left( e^{b x} \textrm{erfc}(ax)   -  e^{\frac{b^2}{2 a^2}} \textrm{erf}\left( \frac{b}{2a}-ax \right)  \right),
\end{align}	
where $\textrm{erf}(\cdot)$ is the error function and $\textrm{erfc}(\cdot)$ is the complementary error function \cite{jeffrey2007table}. 
Applying a change of variables $x=[\ln(\Upsilon) - \ln(\Upsilon_1) + 2C_5]$, and given the fact that $Q(x) = \frac{1}{2} \textrm{erfc}\left(\frac{x}{\sqrt{2}}\right)$ and $Q(x) = \frac{1}{2}- \frac{1}{2} \textrm{erf}\left(\frac{x}{\sqrt{2}}\right)$, the closed form expression for \eqref{a2} is derived in \eqref{a4}.

%%-------------------------------------------------
%%-------------------------------------------------
\section{The Proof of Theorem 2}
\label{strong_dist}
We reformulate \eqref{s} as
\begin{align}
	\label{st1}
	h = \underbrace{h_{l_{ug}} h_{l_{gu}} h_{p_g}}_{h_c}  
	     \underbrace{ h_{a_{gu}} \overbrace{h_{a_{ug}} h_{p_u}}^{h'} }_{h''} 
	     h_{MRR}.
\end{align}
In the sequel, we have used the PDF of product of two RVs, which is generally given as \cite{papoulis2002probability}
\begin{align}
\label{st2}
f_z(z) = \int_{-\infty}^\infty \frac{1}{|x|} f_x(x)  f_y\left(\frac{z}{x}\right)  \textrm{d}x,
\end{align}
where $z$ is the product of RVs $x$ and $y$ ($z = xy$).
We express the $k_\nu(.)$ in terms of the Meijer's G-function as $k_\nu(x)=G^{2,0}_{0,2}\left( \frac{x^2}{4} \bigg|
\begin{array}{c}
	-  \\
	(\nu/2),-(\nu/2)  
\end{array}
\right)$ \cite{jeffrey2007table}. Substituting \eqref{ss8} and \eqref{sr6} in \eqref{st2} and using \cite[eqs. (07.34.21.0002.01) and (07.34.17.0007.01)]{wolfram}, the PDF of RV $h'$ is obtained as
\begin{align}
\label{st3}
f_{h'}(h') &= \frac{\pi w_z^2 \alpha \beta K}{2A_r \Gamma(\alpha)\Gamma(\beta)} \\
%-------------
&~~~\times G^{3,0}_{1,3}\left( \frac{\pi w_z^2 \alpha \beta h'}{2A_r} \bigg|
\begin{array}{c}
	K  \\
	K-1,\alpha-1, \beta-1  
\end{array}
\right). \nonumber
\end{align}
Now, substituting \eqref{ss8} and \eqref{st3} in \eqref{st2} and using \cite[eq. (21)]{adamchik1990algorithm}, the PDFof RV $h''=h_{a_{gu}} h'$ is derived in \eqref{st4}.
%
%
%%%%%%%%%%%%%%%%%%%%%%%%%%%%%%%%%
%%%%%%%%%%%%%%%%%%%%%%%%%%%%%%%%%
\begin{figure*}
	\normalsize
	\begin{align}
		\label{st4}
	f_{h''}(h'') = \frac{K  \left(\pi w_z^2 \alpha^2 \beta^2  \right)^{\frac{\alpha+\beta}{2}}}
	{(\Gamma(\alpha)\Gamma(\beta))^2  \left(2A_r  \right)^{\frac{\alpha+\beta}{2}}} 
	h''^{\frac{\alpha+\beta}{2}-1}
	%-------------
	G^{0,5}_{5,1}\left( \frac{2A_r }{ \pi w_z^2\alpha^2 \beta^2 h''} \bigg|
	\begin{array}{c}
		\frac{2-\alpha+\beta}{2}, \frac{2-\beta+\alpha}{2},  \frac{2+\alpha+\beta}{2}-K, 
		\frac{2+\beta-\alpha}{2}, \frac{2+\alpha-\beta}{2}
		\\
		\frac{\alpha+\beta}{2} -K 
	\end{array}
	\right).  
	\end{align}
	\hrulefill
	\vspace*{4pt}
\end{figure*}
%%%%%%%%%%%%%%%%%%%%%%%%%%%%%%%%
%%%%%%%%%%%%%%%%%%%%%%%%%%%%%%%%
%
%
Using \cite[eqs. (07.34.16.0001.01) and (07.34.16.0002.01)]{wolfram}
the \eqref{st4} can be further simplified as \eqref{st5}.
\begin{align}
\label{st5}
&f_{h''}(h'') = \frac{  \pi w_z^2 \alpha^2 \beta^2 K  }
{2A_r (\Gamma(\alpha)\Gamma(\beta))^2 }  \times \\
%-------------
&G^{5,0}_{1,5}\left( \frac{ \pi w_z^2\alpha^2 \beta^2 h''}{2A_r } \bigg|
\begin{array}{c}
	K\\
	\alpha-1, \beta-1, K-1,\alpha-1,\beta-1
\end{array}
\right).  \nonumber
\end{align}

\begin{figure}
	\begin{center}
		\includegraphics[width=3.2 in]{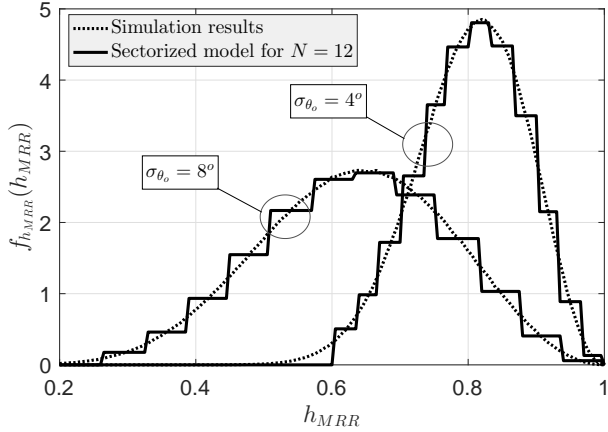}
		\caption{ Sectorized model of the PDF of $h_{MRR}$ for $N=12$ and comparison with the distribution of $h_{MRR}$ obtained using simulation for two different values of $\sigma_{\theta_o}$.  }
		\label{sec_fig}
	\end{center}
\end{figure}

We propose an approximate sectorized model for $f_{h_{MRR}}(h_{MRR})$ as
\begin{align}
\label{st6}
f_{h_{MRR}}&(h_{MRR}) \simeq \\
&\sum_{n=1}^N  B_n \left[ U(h_{MRR} - V_n) - U(h_{MRR} - V_{n+1}) \right], \nonumber
\end{align}
where
$$ V_n= \left\{
\begin{array}{ll}
	&\!\!\!\!\!\!\!2 \mu_{MRR}-1, ~~~~~~~~~~~~~~~~~~~ n=1,\\
	%-------
	&\!\!\!\!\!\!\! V_{n-1}+\frac{(2-2\mu_{MRR})n}{N}, ~~~ n\in\{2,...,N+1\},
\end{array}
\right.$$
and $U(x) = \left\{
\begin{array}{ll}
	&\!\!\!\!\!\!\!1,~~~x>0, \\
	%-------
	&\!\!\!\!\!\!\!0,~~~x<0,
\end{array}
\right. $ is the Heaviside step function. 
Also, the coefficients $B_n$ in \eqref{st6} depend on the parameters $\sigma_{\theta_o}$ and the number of sectors denoted by $N$.
In Fig. \ref{sec_fig}, we compare the proposed sectorized model with respect to the distribution of $h_{MRR}$ obtained using simulation for two different values of $\sigma_{\theta_o}$ and $N=12$.
%
%
%%%%%%%%%%%%%%%%%%%%%%%%%%%%%%%%%%%%%%%%%%%%%%%%%%%%%%%%%%%%%%%%
%%%%%%%%%%%%%%%%%%%%%%%%%%%%%%%%%%%%%%%%%%%%%%%%%%%%%%%%%%%%%%%%
%%%%%%%%%%%%%%%%%%%%%%%%%%%%%%%%%%%%%%%%%%%%%%%%%%%%%%%%%%%%%%%%
%%%%%%%%%%%%%%%%%%%%%%%%%%%%%%%%%%%%%%%%%%%%%%%%%%%%%%%%%%%%%%%%
%
%
Obviously, the accuracy of the proposed sectorized model directly depends on the number of sectors $N$, and for sufficiently large values of $N$, an exact match between simulations and analysis can be achieved at the cost of higher complexity. 
%Hence, choosing an optimal value for $N$ involves a tradeoff between tolerable complexity and desirable accuracy. 
%
Next, we use the proposed model to find a closed-form analytical model for $h$ under moderate to strong turbulence conditions. 
In the section of simulation results, it is shown that $N=8$ achieves a sufficient accuracy.  
In Table \ref{sec_tab}, the coefficients $B_n$ obtained using simulation for different values of $\sigma_{\theta_o}$ and $N=8$.
For the rest of $\sigma_{\theta_o}$, the related coefficients $B_n$ can be obtained by interpolation from the given values.
Substituting \eqref{st5} and \eqref{st6} in \eqref{st2}, we obtain \eqref{st7}.
%
%
%%%%%%%%%%%%%%%%%%%%%%%%%%%%%%%%%
%%%%%%%%%%%%%%%%%%%%%%%%%%%%%%%%%
\begin{figure*}[t]
	\normalsize
	\begin{align}
		\label{st7}
		&f_{h}(h) = \frac{K  \left(\pi w_z^2 \alpha^2 \beta^2  \right)^{\frac{\alpha+\beta}{2}}}
		{(\Gamma(\alpha)\Gamma(\beta))^2  \left(2A_r h_c  \right)^{\frac{\alpha+\beta}{2}}} 
		h^{\frac{\alpha+\beta}{2}-1} \sum_{n=1}^N B_n \nonumber \\
		%
		%-------------
		&\left[\int_0^{V_{n+1}}   h_{MRR}^{-\frac{\alpha+\beta}{2}}
		G^{0,5}_{5,1}\left( \frac{2A_r h_c }{ \pi w_z^2\alpha^2 \beta^2 h}h_{MRR} \bigg|
		\begin{array}{c}
			\frac{2-\alpha+\beta}{2}, \frac{2-\beta+\alpha}{2},  \frac{2+\alpha+\beta}{2}-K, 
			\frac{2+\beta-\alpha}{2}, \frac{2+\alpha-\beta}{2}
			\\
			\frac{\alpha+\beta}{2} -K 
		\end{array}
		\right) \textrm{d} h_{MRR} \right. \nonumber \\
		%--------------
		&-\left.\int_0^{V_{n}}   h_{MRR}^{-\frac{\alpha+\beta}{2}}
		G^{0,5}_{5,1}\left( \frac{2A_r h_c }{ \pi w_z^2\alpha^2 \beta^2 h}h_{MRR} \bigg|
		\begin{array}{c}
			\frac{2-\alpha+\beta}{2}, \frac{2-\beta+\alpha}{2},  \frac{2+\alpha+\beta}{2}-K, 
			\frac{2+\beta-\alpha}{2}, \frac{2+\alpha-\beta}{2}
			\\
			\frac{\alpha+\beta}{2} -K 
		\end{array}
		\right) \textrm{d} h_{MRR} \right]. 
	\end{align}
	\hrulefill
	\vspace*{4pt}
\end{figure*}
%%%%%%%%%%%%%%%%%%%%%%%%%%%%%%%%
%%%%%%%%%%%%%%%%%%%%%%%%%%%%%%%%
%
%
%
Finally using \cite[07.34.21.0084.01]{wolfram} and after some manipulations, the closed-form expression for $h$ is derived in \eqref{st8}.
Substituting \eqref{st8} in \eqref{cdf_general} and using \cite[eq. (07.34.21.0084.01)]{wolfram}, the CDF of $h$ is derived in \eqref{st9}.
Based on \eqref{k1}, we obtain
\begin{align}
	\label{xcc1}
	F_\Upsilon(\upsilon) = F_h\left( \sqrt{\frac{\Upsilon}{\Upsilon_1}}\right),
\end{align}
and then substitute \eqref{cdf_strong} into \eqref{xcc1}, the CDF of $\Upsilon$ can be derived in  \eqref{p2}.

Furthermore, from \eqref{st6}, for lower values of $\sigma_{\theta_o}$, the \eqref{st8} and \eqref{strong_snr_pdf} can be simplified as \eqref{st9} and \eqref{xp1}, respectively.
Now, substituting \eqref{st9} and \eqref{xp1} in \eqref{cdf_general} and using \cite[eq. (07.34.21.0084.01)]{wolfram}, the CDF of $h$ and $\Upsilon$ are approximated as \eqref{cdf_aprox_strong} and \eqref{p3}, respectively.

Also, for small values of $\sigma_{t_o}$ (lower than $0.05^o$), $f_{h_{MRR}}=({h_{MRR}})$ can be well approximated by a Dirac delta function as $f_{h_{MRR}}({h_{MRR}})\simeq\delta(h_{MRR}-1)$. From this, and by following the method used for  obtaining \eqref{st8} and \eqref{cdf_strong}, the PDF  and CDF of $h$ can be approximated as \eqref{st9} and \eqref{cdf_aprox_strong}, respectively.

%we can derive the PDF of h eq in Meijer-G function as:

%----------- end --------------------------------------------
%----------- end --------------------------------------------

%%-------------------------------------------------
%%-------------------------------------------------
\section{The Proof of Proposition 6}
\label{strong_BER}
%%-------------------------------------------------
%%-------------------------------------------------
Lets to rewrite the Gaussian Q-function as complementary error function by $Q(x) = 2\text{erfc}\left(\sqrt{2}x\right)$.
Also, using \cite[eq. (06.27.26.0006.01)]{wolfram}, we can rewrite $Q(x)$ as
\begin{align}
	\label{b2}
	Q(x) = \frac{1}{2\sqrt{\pi}}  G^{2,0}_{1,2}\left(\frac{x^2}{2} \bigg|
	\begin{array}{c}
		1
		\\
		0,1/2
	\end{array}
	\right).
\end{align}
Using \eqref{strong_snr_pdf} and \eqref{a1}, the BER of the considered system can be obtained as 
\begin{align}
	\label{b3}
	&\mathbb{P}_e = \frac{B_s}{4\sqrt{\pi\Upsilon_1 \Upsilon}}\sum_{n=1}^N   B_n
	\int_0^\infty   G^{2,0}_{1,2}\left(\frac{\Upsilon}{2} \bigg|
	\begin{array}{c}
		1
		\\
		0,1/2
	\end{array}
	\right) \nonumber \\
	&\left[
	G^{6,0}_{2,6}\left( B_n' \sqrt{\frac{\Upsilon}{\Upsilon_1}} \bigg|
	\begin{array}{c}
		K,1
		\\
		0,\alpha-1,\beta-1,K-1,\alpha-1,\beta-1 
	\end{array}
	\right)-  \right. \nonumber \\
	%--------------
	&\left.G^{6,0}_{2,6}\left(B_n'' \sqrt{\frac{\Upsilon}{\Upsilon_1}} \bigg|
	\begin{array}{c}
		K,1
		\\
		0,\alpha\!-\!1,\beta\!-\!1,K-1,\alpha-1,\beta-1 
	\end{array}
	\right) \right]	~\textrm{d} \Upsilon .
\end{align}
%%%%%%%%%%%%%%%%%%%%%%%%%%%%%%%%
%%%%%%%%%%%%%%%%%%%%%%%%%%%%%%%%
%
%
Finally, using \eqref{b3} and \cite[eq. (21)]{adamchik1990algorithm}, after some manipulations, the closed-form expressions for BER is derived in \eqref{b5}.


\end{document}